\input harvmac
\input epsf.tex

\def \De {{\Delta}}

\def \ra {\rightarrow}

\def \lra  {\leftrightarrow}

\def \ab{\bar{\alpha}}
\def \four {{1\ov 4}}

\def \ep {\epsilon}

\def \N {{\cal N}}

\def \x {\xi} 
\def \C {{\cal C}}
\def \e {{\eta}}

\def \g {\gamma}
\def \del {\partial}

\def \a {\alpha}
\def \b {\beta}
\def \r {\rho}
\def \s {\sigma}
\def \p {\phi}
\def \m {\mu}
\def \n {\nu}

\def \t {\theta}
\def \Ga {{\Gamma}}
\def \td {\tilde }
\def \d {\delta}

\def \C {{\cal C}}
\def \P {\Phi}

\def \ov {\over }

\def \KK {{\cal K}}

\def \ha {{{1 \over 2}}}

\def \L {{\Lambda}}

\def \l {\lambda}

\def \lr { \lref}
\def\np {{  Nucl. Phys. }}
\def \pl {{  Phys. Lett. }}

\def \pr  {{ Phys. Rev. }}

\def \cmp {{ Commun. Math. Phys. }}

\def \te {{ \td{\ep} }}
\def \tde {{ \td{\d} }}
\def \cft {${{\rm CFT}_d}$}
\def \ads {${\rm AdS}_{d+1}$}

\newcount\figno
\figno=0
\def\fig#1#2#3{
\par\begingroup\parindent=0pt\leftskip=1cm\rightskip=1cm\parindent=0pt
\baselineskip=11pt
\global\advance\figno by 1
\midinsert
\epsfxsize=#3
\centerline{\epsfbox{#2}}
\vskip 0.2cm
{\bf Figure \the\figno:} #1\par
\endinsert\endgroup\par
}
\def\figlabel#1{\xdef#1{\the\figno}}
\def\encadremath#1{\vbox{\hrule\hbox{\vrule\kern8pt\vbox{\kern8pt
\hbox{$\displaystyle #1$}\kern8pt}
\kern8pt\vrule}\hrule}}

\baselineskip 15pt
\Title{\vbox
{\baselineskip 6pt
{\hbox {Imperial/TP/98-99/12}}
{\hbox{hep-th/9811152}} 
{\hbox{   }}
}}
{\vbox{
\centerline {Scattering in Anti-de Sitter Space }
\smallskip
\centerline {and Operator Product Expansion}
 \vskip4pt }}

\centerline  { Hong Liu\footnote {$^*$} 
{e-mail address: hong.liu@ic.ac.uk} 
}

\medskip

\smallskip\smallskip
\centerline {\it  Theoretical Physics Group, Blackett Laboratory,}
\smallskip
\centerline {\it  Imperial College,  London SW7 2BZ, U.K.}
\bigskip\bigskip
\centerline {\bf Abstract}
\medskip
\baselineskip14pt
\noindent
\medskip

We develop a formalism to evaluate generic scalar exchange diagrams in \ads\
relevant for the calculation of four-point functions in AdS/CFT 
correspondence.  
The result  may  be written as an infinite power series of functions of 
cross-ratios. Logarithmic singularities appear in all orders 
whenever the dimensions of involved operators
satisfy certain relations. 
We show that the \ads\ amplitude can be written in a form recognisable
as the conformal partial wave expansion of a four-point function 
in \cft\ and identify the spectrum of intermediate operators. 
We find that, in addition to  the contribution of  the
scalar operator associated with
the exchanged  field in the \ads\ diagram, there are also  
contributions of some other operators which may possibly be 
identified with two-particle bound states in \ads.
The \cft\ interpretation also provides a useful way to ``regularize''
the logarithms appearing in \ads\ amplitude.

\Date {November 1998}

\noblackbox \baselineskip 18pt  


\lr \thooft{G. 't Hooft, ``A Planar diagram theory for strong interactions'',
Nucl. Phys. B72 (1974) 461.}

\lr \pol{ A.M. Polyakov, ``String theory and
quark confinement", hep-th/9711002; 
 ``A few projects in
string theory", Les Houches Summer  School,  1992:783, 
hep-th/9304146. }

\lr \mal{ J. Maldacena, ``The large $N$ limit of
superconformal
field theories and supergravity", {Adv.Theor.Math.Phys. 2 (1998) 231, 
hep-th/9711200}.}

\lr \gkp {S.S. Gubser, I.R.  Klebanov and A.M.  Polyakov,
``Gauge theory correlators from non-critical string theory",
Phys.Lett. B428 (1998) 105, hep-th/9802109.}

\lr \witt {E.  Witten, ``Anti de Sitter space and holography", 
Adv.Theor.Math.Phys. 2 (1998) 253, hep-th/9802150.}

\lr \ferrz{ S. Ferrara,   C. Fronsdal and A. Zaffaroni, 
``On N=8 Supergravity on $AdS_5$ and $N=4$
Superconformal Yang-Mills theory", 
hep-th/9802203.}

\lr \ferza{ S. Ferrara  and A. Zaffaroni, 
``Bulk gauge fields in AdS supergravity and supersingletons",   
hep-th/9807090.}

\lr \ooo{ G. Horowitz and H. Ooguri,
``Spectrum of Large N Gauge Theory from Supergravity",
 Phys. Rev. Lett. 80 (1998) 4116, hep-th/9802116.
 }

\lr \kl { I.R. Klebanov, ``World-volume approach
to absorption by non-dilatonic branes",
\np B496 (1997) 231, hep-th/9702076;
S.S. Gubser, I.R.  Klebanov and A.A.
  Tseytlin, ``String theory and classical
  absorption by three-branes", 
  \np B499 (1997) 41, hep-th/9703040;
S.S. Gubser and I.R. Klebanov,
 ``Absorption by branes and Schwinger
terms in the world-volume
theory", Phys. Lett. B413 (1997) 41, 
hep-th/9708005.}

\lr\hashi{S.S. Gubser, A. Hashimoto, I.R. Klebanov  and  M.
       Krasnitz,
       ``Scalar absorption and the breaking of the world volume conformal
       invariance",
       hep-th/9803023;
 S. S. Gubser and A. Hashimoto, ``Exact absorption probabilities for the D3-brane'',
hep-th/9805140.}

\lr \volo{I.Ya. Aref'eva and  I.V. Volovich,
   ``On large $N$ conformal theories, field theories in Anti de Sitter 
space  and singletons", hep-th/9803028;
``On the Breaking of Conformal Symmetry in the AdS/CFT Correspondence", hep-th/9804182.}

\lr \free{D.Z. Freedman, S.D. Mathur, A. Matusis and L. Rastelli, 
``Correlation functions in the CFT$_d$/AdS$_{d+1}$
correspondence", hep-th/9804058.}

\lr\freed{D.Z. Freedman,  Strings '98 lecture, 
http://www.itp.ucsb.edu/online/strings98/ . }

\lr\hofree{E. D'Hoker and  D. Z. Freedman, ``Gauge Boson Exchange 
in $AdS_{d+1}$'', hep-th/9809179.}

\lr \liuh{H. Liu and A. A. Tseytlin, ``D=4 Super-Yang-Mills, D=5 Gauged
Supergravity, and D=4 Conformal Supergravity,''  
Nucl. Phys. B533 (1998) 88-108, hep-th/9804083.}

\lr \liut{H. Liu and A. A. Tseytlin, ``On Four-point functions in the CFT/AdS
correspondence'', hep-th/9807097.}

\lr\muck{
W.  M\" uck and K.S. Viswanathan, 
``Conformal field theory correlators from classical scalar field theory on AdS$_{d+1}$", Phys.Rev. D58 (1998) 041901, hep-th/9804035.}

\lr \muckv{W.  M\"uck and  K. S. Viswanathan, ``Conformal Field
Theory Correlators from Classical Field Theory on Anti-de Sitter Space II.
Vector and Spinor Fields,'' Phys.Rev. D58 (1998) 106006, hep-th/9805145.}

\lr\nensfet{
M. Henningson and  K. Sfetsos,
``Spinors and the AdS/CFT
correspondence",  hep-th/9803251. }

\lr \rchalmers{G. Chalmers, H. Nastase, K. Schalm and R. Siebelink, 
 ``$R$-Current Correlators in $\N=4$ SYM from $AdS$,''
hep-th/9805015.}

\lr \gordon{G. Chalmers and K. Schalm, ``The large $N_c$ limit of 
Four-point functions in N=4 super Yang-Mills theory
from Anti-de Sitter Supergravity'', hep-th/9810051.}

\lr\brodie{J. H. Brodie and  M. Gutperle, ``String corrections to 
four point functions in the AdS/CFT correspondence'', hep-th/9809067.}

\lr \ramir{A. M. Ghezelbash, K. Kaviani, S. Parvizi and  A. H. Fatollahi,
``Interacting Spinors-Scalars and the AdS/CFT Correspondence,''
hep-th/9805162.}

\lr \sei{S. Lee, S. Minwalla, M. Rangamani and N. Seiberg, 
``Three-Point Functions of Chiral Operators in D=4, $\N=4$ SYM at Large N'',
hep-th/9806074.}

\lr \fro{G.E. Arutyunov and S. Frolov, ``On the origin of supergravity boundary
terms in the AdS/CFT correspondence'', hep-th/9806216.} 

\lr \howe{P.S. Howe and P.C.  West, ``Operator product expansions in 
four-dimensional superconformal field theories'', 
Phys. Lett. B389 (1996) 273, hep-th/9607060.}

\lr \ansel{D. Anselmi, M. Grisaru and A. Johansen, ``A Critical Behaviour 
of Anomalous Currents, Electric-Magnetic Universality and $CFT_4$'', 
hep-th/9601023,
Nucl.Phys. B491 (1997) 221.}

\lr \allen{
B. Allen, ``Graviton propagator in de Sitter space'', \pr D 34 (1986) 3670;
B. Allen and M. Turyn, ``An evaluation of the graviton propagator
in de Sitter space'' \np B292 (1987) 813.}

\lr \fpf{E.S.  Fradkin and M.Ya. Palchik,
   ``Conformal Quantum Field Theory in D
   dimensions"  
   (Kluwer, Dordrecht, 1996).}

\lr \fps{E.S.  Fradkin and M.Ya. Palchik,
   Phys. Rept. 44C (1978) 249.}

\lr \tmp{I.T.  Todorov, M.C. Mintchev and V.B. Petkova
   ``Conformal Invariance in Quantum Field Theory 
   (Scuola Normale Superiore, Pisa, 1978).}

\lr \cpw{
A. M. Polyakov, ``Non-Hamiltonian 
approach to conformal  quantum  field theory'', JETP 39 (1974) 10;
M. Ya. Palchik, ``On the dynamic nature of global
conformal transformations'', \pl 66B (1977) 259; 
G. Mack ``Convergence of operator product expansions on the vacuum
in conformal  invariant quantum  field theory'', \cmp 53 (1977) 155.}

\lr \ferr{S. Ferrara, R. Gatto, A.F. Grillo and G. Parisi,
``Covariant expansion of the conformal four-point function'',
\np B49 (1972) 77; `Analytic properties and asymptotic expansions
of conformal covariant Green's functions'', Nuovo Cimento 19A (1974) 667.}

\lr\ferrg{S. Ferrara and  A. Zaffaroni, ``Bulk Gauge Fields in 
AdS Supergravity and Supersingletons'', hep-th/9807090;
S. Ferrara, R. Gatto and A.F. Grillo,  Nuovo Cimento 2  (1971) 1363.}

\lr\frances{P. Di Francesco, P. Mathieu and D. Senechal, ``Conformal
Field Theory'' (Springer 1997).}

\lr\cole{S. Coleman, ``Aspects of Symmetry'' (Cambridge, 1985).}

\lr \bianchi{ M. Bianchi, M.B. Green, S. Kovacs and 
G. Rossi, ``Instantons in supersymmetric Yang-Mills and D-instantons 
in IIB superstring theory'', JHEP 9808 (1998) 013, hep-th/9807033;
M. Bianchi and S. Kovacs, ``Yang-Mills instantons vs. type IIB D-instantons'', 
hep-th/9811060.}

\lr\prop{C. Fronsdal, Phys. Rev. D10 (1974) 589;
C.P. Burgess  and C. A. Lutken, Propagators and Effective Potentials in
Anti-de Sitter Space'', Nucl. Phys.  B272 (1986) 661;
T.  Inami and H. Ooguri, ``One Loop Effective Potential in Anti-de Sitter 
Space'', Prog. Theo. Phys. 73 (1985) 1051;
C.J. Burges, D.Z. Freedman, S. Davis, and G.W. Gibbons, ``Supersymmetry
in Anti-de Sitter Space'', Ann. Phys. 167 (1986) 285.} 

\lr\ghyp{W.N. Bailey, ``Generalised 
Hypergeometric Series'' (Hafner, New York, 1972); 
L. J. Slater, ``Generalised Hypergeometric 
Functions'' (Cambridge 1966).}

\lr \jacob{B. Allen and T. Jacobson, ``Vector two-point functions
in maximally symmetric spaces'', Commun. Math. Phys. 103 (1986) 669.}

\lr\freedm{D.Z. Freedman, S.D. Mathur, A. Matusis and L. Rastelli, 
``Comments on 4-point functions in the CFT/AdS correspondence",
 hep-th/9808006.}

\lr\freedh{D.Z. Freedman and E. D'Hoker, 
``General Scalar Exchange in AdS$_{d+1}$'', hep-th/9811257.}

\lr\petkou{K. Lang and W. Ruhl, ``The Critical O(N) Sigma Model At 
Dimension $2 < D < 4$ and Order $1/N^2$ Operator Product Expansion and 
Renormalizaztion'', Nucl. Phys. B 377 (1992) 371;
A. C. Petkou, ``Conserved Currents, Consistency Relations and Operator 
Product Expansions in the Conformally Invariant O(N) Vector Model'', 
Ann. Phys. 249 (1996); A. C. Petkou, ``Operator Product Expansions and 
Consistency Relations in a O(N) Invariant Fermionic CFT for $2<d<4$'',
Phys. Lett. B 389 (1996).}

\newsec{Introduction}
  
There has been  a recent revival  of interest  in  the connection between 
large $N$ Yang-Mills theory \refs{\thooft} and  string theory \refs{\pol}  
following the conjecture \refs{\mal} that there is an exact 
correspondence \refs{\gkp,\witt} between 
IIB superstring theory on $AdS_5 \times S_5$
and $\N=4$ Super-Yang-Mills theory in four dimension (see also  \kl).

Under this proposal, correlation functions of $\N=4$ super-Yang-Mills
(SYM) with gauge group $SU(N)$ in the large-$N$ and 
large 't Hooft coupling limit
can be obtained by evaluating  scattering amplitudes of type IIB supergravity
on $AdS_5 \times S_5$. Some   `model' and `realistic'  
two-point and three-point functions have been computed  in 
\refs{\volo, \muck, \nensfet,  \free, \liuh, \rchalmers, \muckv, 
\sei, \ramir}. Since  the structures of two- and three-point functions 
are severely restricted by conformal symmetry, in many cases the 
computations amount to fixing the overall constants.
Four-point functions can be  arbitrary functions of cross-ratios and thus  
encode more dynamical information. Recently some efforts have 
been made in  this direction  \refs{\freed,\liut,\freedm,\hofree,
\brodie,\gordon,\bianchi} aiming to understand 
more about the non-perturbative dynamics of $\N=4$ SYM.

Considering a \cft \foot{Though our prime interest is $\N=4$ SYM,
most of our discussion will apply to any $d-$dimensional conformal field
theory appearing in \ads/\cft\ correspondence.},  we shall assume that 
there exists a closed operator algebra,
which is a strong version of the Wilson operator product 
expansion,
\eqn\wope{
O_{i}(x) O_j(0)  = \sum_k C^k_{ij}(x) O_k(0) \ .
}
Here the summation is over all operators and their coordinate 
derivatives, and $C^k_{ij}$ are c-number functions.
From \wope, a four-point function may be expanded in terms of 
conformal partial waves, e.g., when  $x_{12}, x_{34} \ra 0$,
as an $s-$channel exchange,
\eqn\cpwe{
<O_{i_1}(x_1) O_{i_2}(x_2) O_{i_3}(x_3) O_{i_4}(x_4)>
= \sum_{j} C^j_{i_1 i_2}(x_{12}) C^j_{i_3 i_4}(x_{34}) 
<O_j(x_2) O_j(x_4)>
}
Alternatively, we can also write the four-point function in terms 
of $t-$ or $u-$channel exchanges in the limit  $x_{14}, x_{23} \ra 0$ 
or $x_{13}, x_{24} \ra 0$.
If the algebra \wope\ is complete and associative,  
all channels of exchange are equivalent.

We would like to examine whether a four-point function 
calculated from the scattering amplitude in \ads\ can be written 
in the form of \cpwe\ as we take the corresponding limits in cross-ratios.
A positive answer would be a confirmation of the assumption of 
a closed algebra \wope, which hitherto has been only known to hold 
in two dimensions. And we could further extract important 
non-perturbative information about \cft\ by  identifying 
the spectra of intermediate operators in each channel. 
In the case of $\N=4$ SYM in large $N$ and
large $g^2 N$ limit, knowledge of four-point functions 
would help us answer  questions  such as \refs{\freed,\liut,\freedm}:

1. Does  $\N=4$ SYM in large $N$ and large $g^2 N$ limit have 
a closed algebra \wope?

2. If yes, what is the spectrum of  operators? 
In particular, do chiral operators, which are in one-to-one 
correspondence with IIB supergravity modes on $AdS_5 \times S_5$,
form a complete set? 

In this paper, we shall make some preliminary 
progress in answering these questions. In particular, we shall 
find indications that there are operators in the spectrum which 
correspond to two-particle bound states in \ads. 

\fig{
Exchange and contact diagrams in \ads: $\l$ and $\l_i$ 
are dimensions of the conformal operators corresponding to the 
fields in the diagrams.   
}{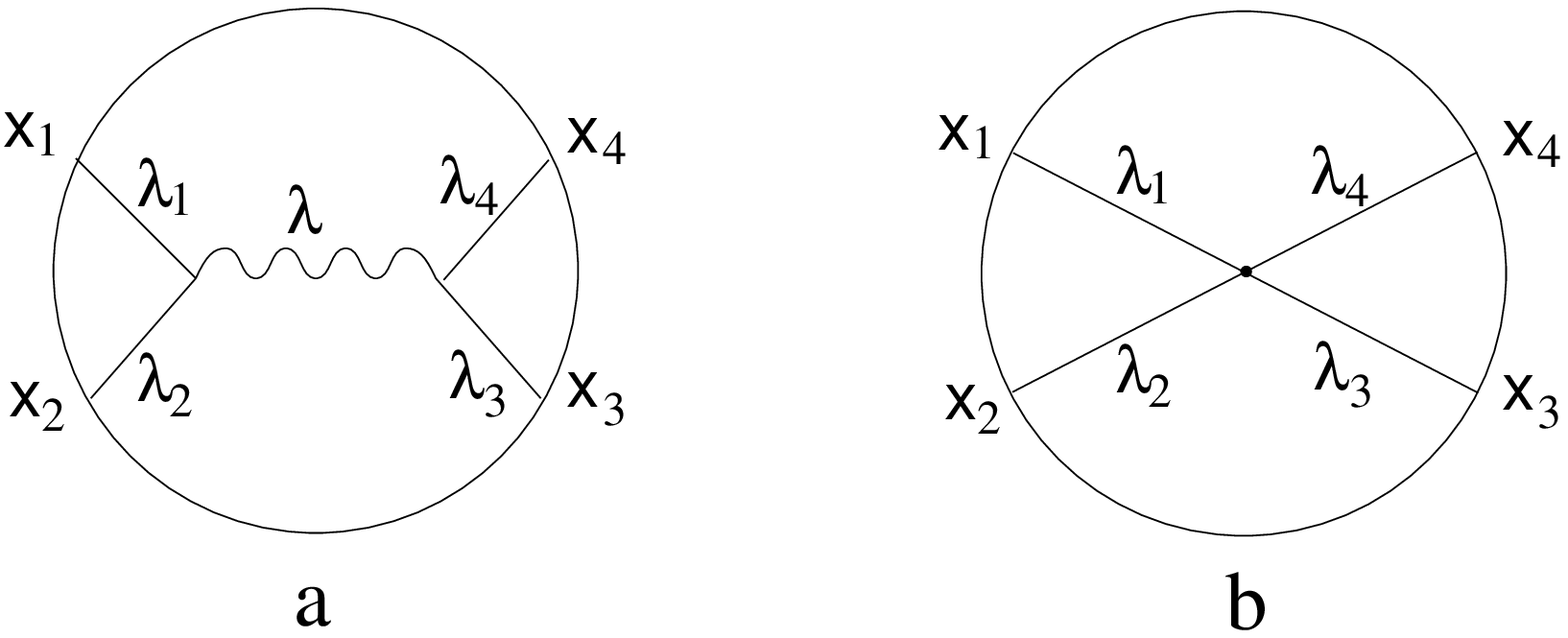}{9.8 truecm}
\figlabel\fcont

One of the obstacles in the computation of realistic four-point
functions in $\N=4$ SYM  has been the difficulty in evaluating 
exchange diagrams \witt\ in AdS space (see figure \fcont a), 
which involve very complicated integrals. 
Here we present a formalism to address this problem (see also
\refs{\hofree}),  providing explicit formulas for AdS integrals  involving 
scalar fields of arbitrary mass needed to evaluate 
generic four-point functions. The  result is written as a single 
inverse Mellin integral so that the analytic properties of the 
amplitudes become transparent. In particular, for the exchange 
diagram of fig. \fcont a, in the limit 
$x_{12}, x_{34} \ra 0$ the scattering amplitude can be written as
a contour integral
\eqn\intmel{
S_{\l} = \int_{\C} \! ds \,\, \Ga ({\l_1 + \l_2 \ov 2} - s) \,
 \Ga ({\l_3 + \l_4 \ov 2} - s) \, \Ga({\l \ov 2} - s) \, H(s, \e, \x) \ .
}
where $\x$ and $\e$ are independent cross-ratios and 
$H$ is a function of complex variable $s$ and $\x,\e$.
In \intmel\ we have only explicitly written down the Gamma functions 
which generate poles inside the contour. $S_{\l}$ can then be
evaluated by the calculus of residues and written as a sum of residues 
of the integrand at three infinite pole sequences (see figure 2). 
We find that  logarithms of cross ratios, 
first found \freedm\ in leading order expansion of some contact 
diagrams, arise generically  whenever the poles in \intmel\
merge into double poles or triple poles, i.e. when
\eqn\logex{
{\l_{1} + \l_{2} - \l_{3}-\l_{4} \ov 2}\,\,{\rm or} \,\,
{\l_{1}+ \l_{2}-\l  \ov 2}\,\,{\rm or} \,\,
{\l_{3}+\l_{4} - \l \ov 2} ={\rm integer} \ .
}
They appear in all orders of the series.
In particular, the contribution from a triple pole 
will contain a part proportional to
$$
(\log|{x_{12} x_{34} \ov x_{13} x_{24}}|)^2  \  .
$$  
Similarly, it can be shown in contact diagrams (see figure \fcont b) 
logarithms occur \freedm\ when
\eqn\logcon{
{\l_{1}+ \l_{2} - \l_{3}- \l_{4} \ov 2} ={\rm integer} \ .
}  

\fig{
Poles and the contour of \intmel. There are three sequences of poles:
1) $s={\l_1 + \l_2 \ov 2} + n$ (represented by solid circles); 
2)$s={\l_3 + \l_4 \ov 2}+n$ (circles);
3) $s={\l \ov 2} + n$ (crosses), where $n=0,1,2,\cdots$.
}{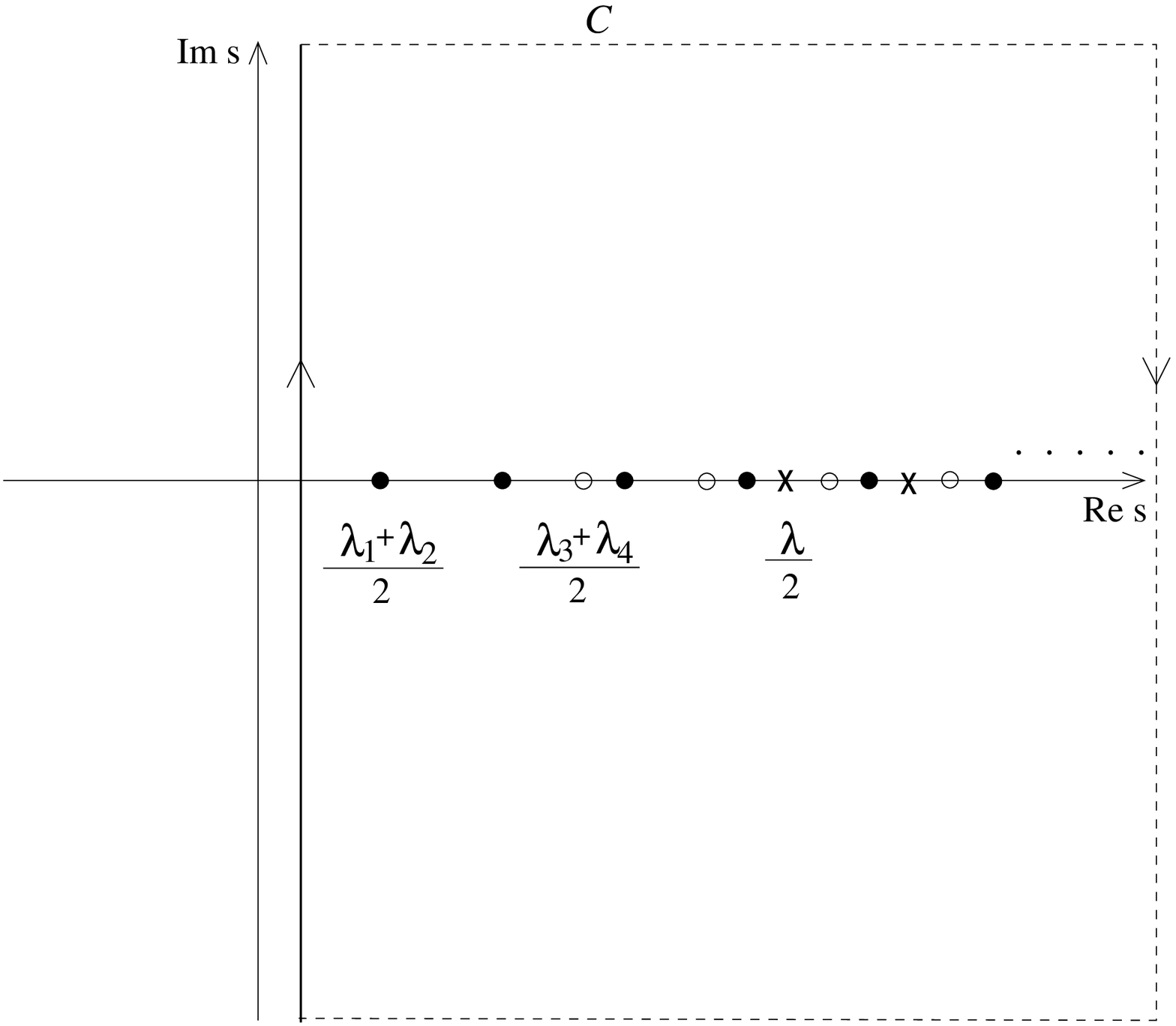}{6.8 truecm}
\figlabel\poleco

In $\N=4$ SYM, 
the dimensions of the chiral fields are protected by supersymmetry
and take integer values. Since in IIB supergravity on $AdS_5 \times S_5$, 
all  vertices are $SU(4)$ singlets, the scattering diagrams 
associated with  four-point functions will in general satisfy \logex\ 
or \logcon. This implies that logarithms are universally present\foot{It 
might still happen that when we add up all the diagrams contributing 
to a four-point function, logarithms will cancel.}.

We then proceed to investigate whether the amplitude 
we find  can be written as the conformal partial wave 
expansion (CPWE) \cpwe. 
The Mellin integral representation \intmel, 
in which our results are presented, turns out to be particularly 
convenient to identify them with $s-$channel OPE exchanges in \cft.
The contribution of each pole sequence 
in \intmel\ can be identified with the CPWE \cpwe\ of 
a conformal operator: the value of a pole corresponds to the scale
dimension of a spin-$0$ descendant\foot{Here we mean an $SO(d,2)$ 
descendant. A spin-$0$ descendant takes the form $(\del^2)^n O$, where
$O$ is the primary and $\del^2$ is the Laplacian.} (we shall call it 
a sub-primary),  while the residue at the pole may be identified 
with the CPWE contribution of a 
subset of descendants associated with the sub-primary.
The pattern may be presented diagrammatically as 
\fig{
$s-$channel OPE interpretation of an exchange diagram. 
}{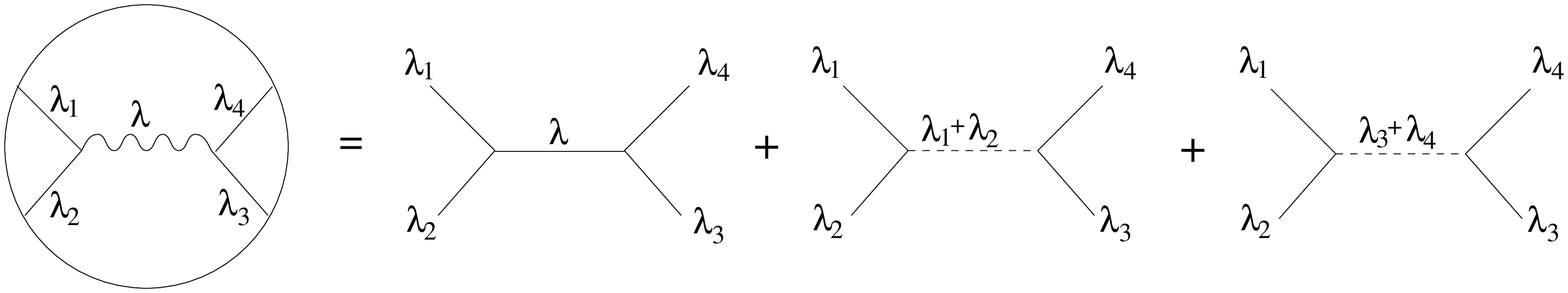}{15.8 truecm}
\figlabel\cftin

The first diagram on the right hand side 
corresponds to the exchange 
of a scalar primary operator of dimension $\l$, which may be interpreted 
as the operator (we shall call it $O_\l$) related to the exchanged 
field in \ads\ by AdS/CFT correspondence. This result was expected earlier
in \liut\ on the basis of indirect considerations. Here we identify the 
contributions of all the descendents of $O_\l$ and show that their 
relative OPE couplings 
\wope\ are  consistent with those required by conformal symmetry. 
The second and third diagram on the right hand side correspond to
the exchanges of operators of dimensions $\l_1 + \l_2 $ and $\l_3
+ \l_4$ respectively (which we shall call $O_{12}$ and $O_{34}$). 
However, in these cases, there are some mismatches in the 
identifications.   
In fig. \cftin\ we have used dotted lines in intermediate states 
to distinguish them from the first diagram.  
Although we have found  contributions from operators having the 
same quantum numbers as  the complete set of descendants of a 
primary operator of dimension $\l_1 + \l_2$ (and $\l_3 + \l_4$), 
the relative OPE couplings \wope\ between the primary and descendants 
seem to be inconsistent with those required by conformal 
symmetry\foot{In a \cft, the OPE coupling \wope\ 
of a descendant is uniquely determined by that of the primary.}.
The OPE couplings of these descendant operators have a peculiar pattern
suggesting the mismatch may be due to some mixing among different 
operators. But we have not been able to make it precise in this paper.

\fig{
$s-$channel OPE interpretation of a contact diagram. 
}{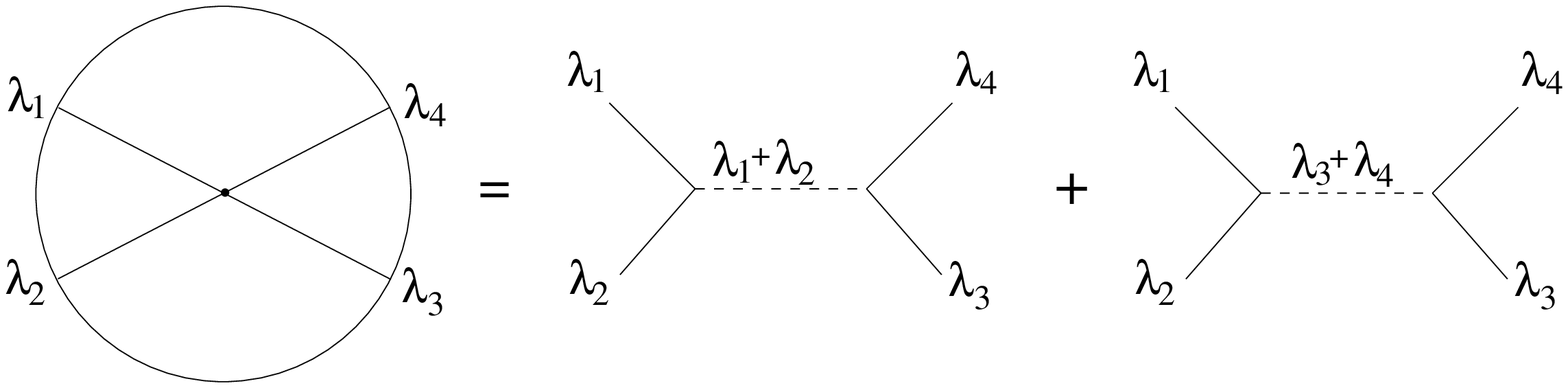}{12.8 truecm}
\figlabel\cftico

Similarly the contact diagram fig. \fcont b may be represented in terms of 
$s-$channel exchanges as in figure \cftico\ and the identifications 
are also not complete in the sense as described in exchange case.

The identification of \ads\ diagrams with CPWE also sheds light on the 
appearance of logarithms. The conditions \logex\ and \logcon\ 
are satisfied precisely when 
the quantum numbers (spin, scale dimensions, etc.) of certain  
descendants  of $O_\l$ or $O_{12}$ or $O_{34}$ become 
identical to one another. 
The OPE couplings in fig. \cftin\ and \cftico\ determined
from \intmel\ fall into the following pattern:
when the quantum numbers of  descendants of different
operators are degenerate, their contributions to the  
conformal partial wave expansion become 
identical and cancel one another. The results are given by their derivatives, 
which contain logarithms. Thus by moving infinitesimally away from the 
degeneracy points \logex\ and \logcon\ in parameter space,  we see that 
the relations in fig.  \cftin\ and \cftico\ provide a physically 
meaningful way to ``regularize'' logarithms.

Although our present analysis based on generic diagrams would not give 
a conclusive answer to the questions  listed earlier, it nevertheless 
provides a starting point for further study.
The relation we find here between 
an arbitrary exchange diagram in \ads\ and CPWE  appears to be {\it universal}
and should be helpful for understanding the general structure of AdS/CFT 
correspondence.
Before having a complete calculation of realistic four-point functions 
in a specific theory, it is probably  premature to speculate about
the relevance of operators $O_{12}$ and $O_{34}$ and the mismatches
in their \cft\ identification.
However, if their contributions are indeed present 
in a realistic amplitude, it should  imply the existence of new operators 
in the spectrum not seen in the Lagrangian of supergravity. 
In $\N=4$ SYM they may be written  as double-trace operators, 
while in AdS$_5$ supergravity they may  be identified with
two-particle bound states.

The plan of the paper is as follows. 
In section two and three we discuss the evaluation of scattering
diagrams in \ads. In section four  we review, for the convenience 
of comparing with \ads\ results, the conformal partial wave expansion 
in \cft. In section five we discuss the 
\cft\ interpretation of \ads\ amplitude.
We have included a number of appendices. In Appendix A we describe 
briefly the subtleties in evaluation of integrals using Mellin transform
and analytic continuation. Appendices B,C are devoted to 
detailed evaluation of some integrals in the main text.

\newsec{Scalar  exchange in anti-de Sitter space}

We consider tree-level scattering  of 
four scalar fields  in \ads\ with masses $m_i, i=1,\cdots,4$
by exchanging a scalar field of mass $m$.
According to AdS/CFT correspondence, a scalar field $\p_i$ of mass 
$m_i$ in \ads\ corresponds to a scalar operator 
$\Phi_{\l_i}$ in \cft, with conformal dimension $\l_i = {d \ov 2} + 
\sqrt{m_i^2 + {d^2 \ov 4}} = {d \ov 2} + \n_i $. The scattering amplitude 
describes in \cft\ the contribution of  
 $\P_{\l}$  to the four-point function of 
scalar operators $\P_{\l_i},  i=1, \cdots, 4$.

In this section, we shall take the interacting vertices 
to be of the form 
$$ 
{\cal L } =\p_1 \, \p_2 \, \p \,\, + \, \, \p_3 \, \p_4 \, \p  \ .
$$ 
Scattering amplitudes resulting from more complicated vertices 
involving derivatives and  contact vertices will be discussed 
in next section.

As in \witt\ we  use the Euclidean (half-space) metric, 
\eqn\adsm{
ds^{2} = g_{\m\n} du^\m du^\n= {1 \over u^2_0} (d u^2_0  + d u_{i}^{2}), 
\  \; \; \; \; \; \; \;       i=1,2, \cdots, d \ .
}
The AdS$_{d+1}$  bulk  indices  will be 
denoted by $\m,\n, ...$ and will take values $ 0, 1,...,  d$.
The points in the bulk are labelled by $u,v, \cdots$, while those
on the boundary by $x,y, \cdots$.
We also use shorthand notations $u= (u_0, \vec{u})$, $x=(\vec{x})$ and
$x_{ij}^2 = |\vec{x}_i - \vec{x}_j|^2$, 
$|u-x_i|^2 = u_0^2 + |\vec{u} -\vec{x}_i|^2$.

The scattering amplitude can then be written as
\eqn\fourt{
S_\l (x_1,x_2,x_3,x_4)=\int \! {d u_0 d^d u \over u_0^{d+1}}
{d v_0 d^d v \over v_0^{d+1}} \ 
\KK_{\l_1} (u,x_1) \KK_{\l_2} (u,x_2)  G (u,v)
\KK_{\l_3} (v,x_3) \KK_{\l_4} (v,x_4)\ , 
}
where $ \KK_{\l_i} (u,x_i), i=1, \cdots, 4$ is the bulk-to-boundary 
propagator \witt\ for field $\p_i$, 
\eqn\bpro{
\KK_{\l_i} (u,x_i)  = c_{\l_i} ({u_0 \over |u-x_i|^2})^{\l_i}\ ,
\,\,\,\,\,\, 
c_{\l_i} = {\Ga (\l_i) \ov \pi^{d \ov 2} \Ga(\n_i)}
}
and  $G(u,v)$ is the AdS bulk scalar propagator \refs{\prop},
\eqn\iui{
G(u,v) =r t^{- \l } F(\l, \n + \ha ; 2 \n +1, t^{-1})
\ . }
In \iui\ $F$  is a hypergeometric function and 
$$
r={\Ga (\l) \ov 2^{2 \l + 1} \pi^{d \ov 2}} {1 \ov \Ga(\n + 1 )},
\,\,\,\,\,\,\,\,\,\,\,\,
t={(u_0 + v_0)^2 + (\vec{u} - \vec{v})^2 \ov 4 u_0 v_0}.
$$

To evaluate \fourt, first we would like to get rid of the 
cross term of $u_0$ and $v_0$ in $t$ in \iui, which complicates 
the integrals. 
This can be achieved by a quadratic
transformation\foot{The one we use here is:
$$
F(a,b;2b;z) = (1-{z \ov 2})^{-a} F (\ha a, \ha (a+1); b+ \ha;
z^2 (2-z)^{-2})
$$}
of the hypergeometric function 
in \iui, after which the bulk propagator becomes, 
\eqn\scap{
G(u,v) = {\Ga (\l) \ov 2^{\l + 1} \pi^{d \ov 2}} {1 \ov \Ga(\n + 1
)} q^{-\l} F({\l +1 \ov 2}, {\l \ov 2};\n+1;{1 \ov q^2}), \,\,\,
}
where
$$
q= {u_0^2 + v_0^2 + |\vec{u}-\vec{v}|^2  \over 2 u_0 v_0}.
$$

Now we use the Mellin-Barnes representation of a hypergeometric function 
\eqn\meba{
F(a,b;c;z) = {\Ga(c) \ov \Ga(a) \Ga(b)} 
 {1 \ov 2 \pi i } \int_{-i \infty}^{i \infty} \! ds 
{\Ga(a+s) \Ga(b+s) \ov \Ga(c+s)} \Ga(-s) (-z)^s
}
in \scap\ and plug it into \fourt. This gives us, 
\eqn\flag{
S_{\l} = C_1 {1 \ov 2 \pi i }
\int_{-i \infty}^{i \infty} \! ds \, {\Ga({\l +1 \ov 2}+s) \Ga({\l \ov 2}+s)
\ov \Ga(\n +1 +s)} \Ga(-s) (-1)^s J(s)
}
with 
$$
J(s) = \int \! {d u_0 d^d u \over u_0^{d+1}}
{d v_0 d^d v \over v_0^{d+1}} \ 
({u_0 \over |u-x_1|^2})^{\l_1} ({u_0 \over |u-x_2|^2})^{\l_2}
({2 u_0 v_0 \ov u_0^2 + v_0^2 + |\vec{u}-\vec{v}|^2 })^{\l + 2s}
$$
\eqn\jflag{
\times \, \, 
({v_0 \over |v-x_3|^2})^{\l_3} ({v_0 \over |v-x_4|^2})^{\l_4}
}
and $C_1 = {1 \ov 4 \pi^{d+1 \ov 2}} \Pi_{i=1}^4 c_{\l_i}$

$J(s)$ still involves quite complicated integrals. 
We present its detailed evaluation in  Appendix B. 
The result can be written in terms of the cross ratios of 
the boundary points as an inverse Mellin type 
integral (for  notations see Appendix B),
$$
J(s) =C_2 \, {1 \ov 2 \pi i}
\int_{{\cal C}}
\! d s_1 \,  \x^{-s_1 - {\De_{34} \ov 2}} \, 
\Ga({\l_{12} \ov 2} -s_1) \Ga ({\l_{34} \ov 2} - s_1) \,
F( {\De_{34} \ov 2} +s_1 ,{\De_{12} \ov 2} + s_1; 2s_1; 1-{\e \ov \x})\,
$$
\eqn\jans{
\times \, \, 
{\Ga ({\l \ov 2} +  s-s_1 ) 
 \ov \Ga ({\l_{12} + \te_{34} \over 2} + s-s_1)} \,
{\Ga( {\De_{34} \ov 2} + s_1) \Ga({\De_{43} \ov 2} + s_1)
\Ga( {\De_{12} \ov 2} + s_1) \Ga({\De_{21} \ov 2} + s_1) 
\ov  \Ga (2s_1)} \
}
where $\e$, $\x$ are cross ratios defined by,   
\eqn\crosr{
\eta = {|x_{13}|^2 |x_{24}|^2
\ov  |x_{12}|^2 |x_{34}|^2}, \,\,\,\,\,\,
\x = {|x_{14}|^2 |x_{23}|^2
\ov  |x_{12}|^2 |x_{34}|^2},
}
and 
$$ 
C_2 = {\pi^d \ov 4} 
{ \Ga ({\l_{12} + \l_{34} -d \ov 2}) 
\Ga (s + {\te_{12} \ov 2}) \Ga (s + {\te_{34} \ov 2})
\ov \Ga(\l_1)\Ga(\l_2) \Ga(\l_3) \Ga(\l_4)
\Ga(\l + 2s)}
{2^{\l + 2s} \ov  |x_{12}|^{\l_{12} + \De_{34}}
|x_{14}|^{\De_{12}-\De_{34}}  
|x_{24}|^{\De_{21}- \De_{34}}
|x_{34}|^{2 \l_3}}
$$

The path of integration ${\cal C}$ in \jans\ (see 
the last paragraph of Appendix B for a more precise description) 
is taken to be parallel to the imaginary $s_1$-axis 
and is deformed if necessary to separate the poles of ascending sequences 
(e.g. those of $\Ga({\l_{12}\ov 2} - s_1)$) 
from the poles of descending sequences (e.g. those of 
$\Ga({\De_{12}\ov 2} + s_1)$) of the integrand.

Plugging the expression for $J$ into \flag, using the duplication formula  
for Gamma functions,
$$
\Ga (\l + 2s) = {1 \ov 2 \pi^{\ha}} 2^{\l + 2s} 
\Ga({\l +1 \ov 2}+s) \Ga({\l \ov 2}+s)
$$ 
and regrouping the terms in the integrand, we find,
$$
S_\l = C_3 \, {1 \ov 2 \pi i}
\int_{{\cal C}}
\! d s_1 \,  \x^{-s_1 - {\De_{34} \ov 2}} \, 
\Ga({\l_{12} \ov 2} -s_1) \Ga ({\l_{34} \ov 2} - s_1) \,
F( {\De_{34} \ov 2} +s_1 ,{\De_{12} \ov 2} + s_1; 2s_1; 1-{\e \ov \x}) \,
$$
\eqn\sef{
\times \,\, 
{\Ga( {\De_{34} \ov 2} + s_1) \Ga({\De_{43} \ov 2} + s_1)
\Ga( {\De_{12} \ov 2} + s_1) \Ga({\De_{21} \ov 2} + s_1) 
\ov  \Ga (2s_1)} 
\,\, I_1 
}
with
\eqn\genhy{
I_1 ={1 \ov 2 \pi i} \int_{-i \infty}^{i \infty} \! d s \, 
\Ga(-s) (-1)^s { \Ga({\te_{12} \over 2} + s)
\Ga({\te_{34} \over 2} + s) \Ga ({\l \over 2} + s-s_1) \ov 
\Ga (s-s_1 + {\l_{12} + \te_{34} \over 2}) \Ga (\n +1 +s)}
}

$I_1$ is nothing but the Mellin-Barnes representation of the generalised
hypergeometric function ${_3}F_2$ \refs{\ghyp}, which leads 
to\foot{If necessary, the 
integration path in \genhy\ should be deformed to separate the poles $s=0,1,
\cdots$ from those poles in descending series.},
\eqn\hyper{
I_1 = { \Ga ({\te_{12} \over 2}) \Ga ({\te_{34} \over 2}) \ov \Ga(\n +1) }
{ \Ga ({\l \ov 2} -s_1) \ov \Ga (-s_1 +  {\l_{12} + \te_{34} \over 2})} \,\,
{_3}F_{2} ({\te_{12} \over 2},{\te_{34} \over 2},{\l \ov 2}-s_1;
{\l_{12} + \te_{34} \over 2}-s_1, \n+1;1)
}
When the parameters of a  
generalised hypergeometric function  ${_3}F_2 (a,b,c;e,f;z)$
satisfy the relation
$e+f = a+b+c +1$, the series will be said to be 
{\it Saalschutzian}\foot{The hypergeometric series ${_3}F_2
(a,b,c;e,f;z)$ converges when $|z|<1$, also when $z=1$ provided that 
$Re(e+f-a-b-c)>0$. Thus we see a {\it Saalschutzian} series is 
convergent at $z=1$. }.
It is easy to check that the hypergeometric series in \hyper\ 
is {\it Saalschutzian}. 
Saalschutz's theorem states that ${_3}F_2 (a,b,c;e,f;z)$ satisfies,
\eqn\saal{
{_3}F_2 (a,b,c;e,f;1) = { \Ga (e) \Ga (1+a-f) \Ga (1+b-f) \Ga (1+c-f) \ov
\Ga (1-f) \Ga (e-a) \Ga (e-b) \Ga (e-c)}
}
provided  $e+f = a+b+c +1$ and  $a,b$ or $c$ is a negative integer.

From eqs \sef\ and \hyper, we reach the final expression for 
$S_\l$, 
$$
S_\l = C \, {1 \ov 2 \pi i} \int_{{\cal C}}
\! d s_1 \, \x^{-s_1} \, \Ga({\l_{12} \ov 2} -s_1) \Ga({\l_{34}\ov 2} - s_1)
 \Ga ({\l \ov 2}-s_1) \, 
F( {\De_{34} \ov 2} +s_1 ,{\De_{12} \ov 2} + s_1; 2s_1; 1-{\e \ov \x})
$$
\eqn\final{
\times \,\, 
{\Ga( {\De_{34} \ov 2} + s_1) \Ga({\De_{43} \ov 2} + s_1)
\Ga( {\De_{12} \ov 2} + s_1) \Ga({\De_{21} \ov 2} + s_1) 
\ov \Ga ({\l_{12} + \te_{34} \over 2} -s_1 ) \Ga (2s_1)}
\,
{_3}F_{2} ({\te_{12} \over 2},{\te_{34} \over 2},
{\l \ov 2}-s_1; {\l_{12} + \te_{34} \over 2} -s_1, \n+1;1)
}
with 
$$
C ={1 \ov 8 \pi^{{3 \ov 2} d}} 
{\Ga ({\l_{34}+\l_{12} -d \ov 2}) 
\Ga ({\te_{12} \over 2}) \Ga ({\te_{34} \over 2})
\ov  \Ga(\n_1) \Ga(\n_2)  \Ga(\n_3)\ \Ga(\n_4) \Ga(\n +1) } 
{1 \ov |x_{12}|^{\l_{12}} \, |x_{14}|^{\De_{12}} \,
|x_{24}|^{\De_{21} - \De_{34}} \, |x_{23}|^{\De_{34}} \, 
|x_{34}|^{ \l_{34}}}
$$

Thus we have been able to reduce \fourt\ to a single inverse Mellin type of 
integral in \final, which may be evaluated by  choosing the appropriate 
contour in the complex $s_1$ plane and the calculus of residues.

Let us consider the $s-$channel OPE limit where $x_{12}, x_{34}$ 
are much smaller  than other distances, 
i.e., $\x, \e \gg 1$ and $1-{\e \ov \x} \ll 1$.
In this case, we can take  the integration path $\C$  over a contour 
enclosing the right half plane and the integral is given 
by the sum of the residues of the integrand at the poles of 
ascending sequences.

On right half plane we have three pole series which come from 
$\Ga ({\l_{12} \ov 2} -s_1)$, $\Ga({\l_{34}\ov 2} - s_1)$ and
$\Ga ({\l \ov 2} -s_1)$ respectively,
\eqn\polese{\eqalign{
& (1) \,\,\, \, s_1 = {\l \ov 2} + n, \,\,\,\,\,\, m=0,1,2, \cdots ; \cr
& (2) \,\,\, \, s_1 = {\l_{12} \ov 2} + n, \,\,\,\,\,\, m=0,1,2, \cdots;\cr
& (3) \,\,\, \, s_1 = {\l_{34} \ov 2} + n,  \,\,\,\,\,\, m=0,1,2, \cdots. 
}}
We first consider the case that no pole series in above coincide
with one another, i.e., none of ${\ep_{12} \ov 2}, \, {\ep_{34} \ov 2}$ and
${\l_{12}-\l_{34} \ov 2}$ is an integer. Then we can write $S_{\l}$ 
as,
$$
S_{\l} = \sum_{n=0}^{\infty} S_n^{(\l)} +
\sum_{n=0}^{\infty} S_n^{(\l_{12})}  + \sum_{n=0}^{\infty} 
S_n^{(\l_{34})}  
$$
where $S^{(\l)}_n$, $S_n^{(\l_{12})}$ and $S_n^{(\l_{34})}$  
are the contributions from the n-th pole in each series.

\subsec{series 1}

Let us first look at the pole series 
$s_1 = {\l \ov 2} +n$. In this case the third parameter
in ${_3}F_{2} ({\te_{12} \over 2},{\te_{34} \over 2}, -s + {\l \ov 2};
-s +  {\l_{12} + \te_{34} \over 2}, \n+1;1)$ becomes a negative 
integer and we can
use  the  Saalschutz's theorem \saal\ to get,
\eqn\sallz{\eqalign{
& {_3}F_{2} ({\te_{12} \over 2},{\te_{34} \over 2},-n;
-s +  {\l_{12} + \te_{34} \over 2}, \n+1;1) \cr
& = {\Ga ({ \l_{12} + \l_{34} -d \ov 2} -n) \Ga ({\ep_{12} \ov 2})
\Ga ({\ep_{34} \ov 2}) \Ga (-n -\n ) 
\ov  \Ga ( -\n )\Ga ({\ep_{12} \ov 2} - n)
\Ga ({\ep_{34} \ov 2} -n )\Ga ({ \l_{12} + \l_{34} -d \ov 2})} \cr
& = (-1)^n { \Ga (1 + \n) \Ga ({\ep_{12} \ov 2}) \Ga ({\ep_{34} \ov 2})
\ov \Ga ({ \l_{12} + \l_{34} -d \ov 2})}
{\Ga ({ \l_{12} + \l_{34} -d \ov 2} -n) \ov 
\Ga ({\ep_{12} \ov 2} - n)
\Ga ({\ep_{34} \ov 2} -n )
\Ga (1+n +\n)}.
}}
where in the second identity we have used the relation:
$\Ga (x) \Ga (1-x) = { \pi \ov sin \pi x}$.

Now plugging \sallz\ into \final, we get 
\eqn\bba{
S_n^{(\l)} = A^{(\l)} \,\, {\x^{-n} \ov n! } 
{\Ga ({\d_1 \ov 2}+ n) \Ga ({\d_2 \ov 2}+ n)
\Ga ({\d_3 \ov 2}+ n) \Ga ({\d_4 \ov 2}+ n)
\ov \Ga(\l + 2n) \Ga (\n +n +1)}
F({\d_3 \ov 2}+ n, {\d_1 \ov 2} + n; \l + 2n; 1-{\e \ov \x}) 
}
with 
$$ 
A^{(\l)} = 
{1 \ov 8 \pi^{{3 \ov 2} d}} 
{\Ga({\ep_{12} \ov 2})\Ga({\ep_{34} \ov 2})
\Ga({\te_{12} \ov 2})\Ga({\te_{34} \ov 2})
\ov  \Ga(\n_1) \Ga(\n_2)  \Ga(\n_3)\ \Ga(\n_4)}
{1 \ov  |x_{12}|^{\ep_{12}} \, |x_{14}|^{ \d_1} \,
|x_{24}|^{\De_{21} - \De_{34}} \, |x_{23}|^{\d_3} \, 
|x_{34}|^{ \ep_{34}}} 
$$
Note that $\Ga(-s)$ has residue $(-1)^{n-1}/n!$ at its pole $s=n$.

\subsec{series 2}

In this case we have $s_1 = {\l_{12} \ov 2} + n$, and 
$$
S_n^{(\l_{12})} = A^{(\l_{12})} \, \, {(-1)^{n} \ov n!} 
\x^{-n} \,\,G_n \,\,
F({\l_{12} + \De_{34} \ov 2}+ n, \l_1 + n; \l_{12} + 2n; 1-{\e \ov \x})
\, \times
$$
\eqn\aaax{
\Ga ({\l_{34}-\l_{12} \ov 2} -n)\,
{\Ga ({\l_{12} + \De_{34} \ov 2}+ n) 
\Ga ({\l_{12} - \De_{34} \ov 2}+ n)
\Ga (\l_1 + n) \Ga (\l_2 + n)  \ov \Ga(\l_{12} + 2n)}
\,\, }
with 
\eqn\coefa{
A^{(\l_{12})} =  {1 \ov 8 \pi^{{3 \ov 2} d}} 
{\Ga ({\l_{34}+\l_{12} -d \ov 2}) 
\ov  \Ga(\n_1) \Ga(\n_2)  \Ga(\n_3)\ \Ga(\n_4)}
{1 \ov  |x_{14}|^{2 \l_1} \,
|x_{24}|^{\De_{21} - \De_{34}} \, |x_{23}|^{\l_{12} + \De_{34}} \, 
|x_{34}|^{ \l_{34}- \l_{12}}}
}
and 
\eqn\ggn{\eqalign{
G_n & = 
{\Ga ({\te_{12} \ov 2})\Ga ({\te_{34} \ov 2}) 
\Ga (-{\ep_{12} \ov 2}-n) \ov \Ga({\te_{34} \ov 2} -n)
\Ga (\n +1)} \,\,
{_3}F_{2} ({\te_{12} \over 2},{\te_{34} \over 2},-{\ep_{12} \ov 2}-n;
{ \te_{34} \over 2} -n, \n+1;1) \cr
& = \sum_{m=0}^{\infty} {1 \ov m !}{\Ga ({\te_{12} \ov 2} + m )\, 
\Ga ({\te_{34} \ov 2} + m) \,
\Ga ( m -{\ep_{12} \ov 2}-n) \ov \Ga( {\te_{34} \ov 2} -n + m)
\Ga (\n +1+m)} 
}}
By a transformation of  $_{3}F_2$ \refs{\ghyp}, \ggn\  can be written in a form
symmetric under $\l \ra  \td{\l}$ and given by a terminating series,
\eqn\ggnt{\eqalign{
G_n & = - {1 \ov {\te_{12} \ov 2} {\ep_{12} \ov 2}} \,
{_3}F_{2}({\l_{12} + \l_{34} -d \ov 2},1,-n;1+{\te_{12} \ov 2},
1+{\ep_{12} \ov 2};1) \cr
& = - { \Ga({\te_{12} \ov 2}) \, \Ga ({\ep_{12} \ov 2}) \ov
\Ga ({\l_{12} + \l_{34} -d \ov 2})} \sum_{m=0}^{n}
(-1)^m {n ! \ov (n-m) !} \, 
{\Ga ({\l_{12} + \l_{34} -d \ov 2} + m) \ov
\Ga(1 + {\te_{12} \ov 2}+ m )  \Ga (1 + {\ep_{12} \ov 2} + m)} \ .
}}
The contribution from poles in series (3) can be obtained from \aaax\ -
\ggn\ by exchanging $1,2$ and $3,4$.

\subsec{Coinciding poles}

When ${\ep_{12} \ov 2}$,  ${\ep_{34} \ov 2}$ or
${\l_{12}-\l_{34} \ov 2}$  become integers, 
the poles from different series  
in \polese\ may merge into double  or triple poles.
For example, when  ${\l_{12}-\l_{34} \ov 2}$ is an integer, 
apart from a finite number of them,  
all  poles in  series two and three in \polese\ will merge into double poles, 
while the poles in the first series remain untouched. 
The contribution from a double pole is given by the derivative  of the 
integrand  of \final. The expressions are quite complicated and we do not 
explicitly write them down here. We simply note that  
there will be terms proportional
to $\ln \x$ as a result of  $\del \x^{-s}/ \del s = - \log \x \, \x^{-s}$.
If all three parameters are integers, 
then apart from a finite number
of simple and double poles all poles may merge into triple poles and their 
contributions
are given by the second derivative of the integrand of \final. 
In these cases, among other things, we will have terms proportional to 
$(\log \x)^2$ from the second derivative of $\x^{-s}$. 

We caution that in certain range of parameters, 
${_3}F_2$ in \final\ may develop zeros at the poles and the pole structure 
may be different from what we naively read from \final. 
This happens when ${\ep_{12} \ov 2}$ or  ${\ep_{34} \ov 2}$ is a 
positive integer.\foot{The following discussion is partly motivated by the 
results in \refs{\freedh}, where simplifications in some expressions 
in this range of parameters
have been observed.  I would like to thank D. Freedman for correspondence
regarding this issue.} 
As an example let us take ${\ep_{12} \ov 2} = k+1 $ 
with $k \geq 0$  an integer. By a transformation \ghyp\ of  
generalized hypergeometric functions, ${_3}F_2$ in eq \final\ can be 
rewritten as,
\eqn\tra{\eqalign{
& {_3}F_{2} ({\te_{12} \over 2},{\te_{34} \over 2},
{\l \ov 2}-s_1; {\l_{12} + \te_{34} \over 2} -s_1, \n+1;1) \cr
& = {\Ga({\l_{12} + \te_{34} \ov 2} -s_1) \ov \Ga({\l_{12} \ov 2} -s_1)
\Ga( 1+ {\te_{34} \ov 2})} {_3}F_{2} ({\te_{34} \over 2}, 1+{\l-d \ov 2}+s_1,
1-{\ep_{12} \ov 2}; 1+ {\te_{34} \over 2}, \n+1;1) \ .
}}
Since $1-{\ep_{12} \ov 2} = -k$, 
${_3}F_{2} ({\te_{34} \over 2}, 1+{\l-d \ov 2}+s_1,
1-{\ep_{12} \ov 2}; 1+ {\te_{34} \over 2}, \n+1;1)$ on the right hand side of
\tra\  is given by a terminating 
series,
\eqn\traa{\eqalign{
&  {_3}F_{2} ({\te_{34} \over 2}, 1+{\l-d \ov 2}+s_1,
1-{\ep_{12} \ov 2}; 1+ {\te_{34} \over 2}, \n+1;1) \cr
& = {_3}F_{2} ({\te_{34} \over 2}, 1+{\l-d \ov 2}+s_1,
-k; 1+ {\te_{34} \over 2}, \n+1;1) \cr
& = {\te_{34} \ov 2} {\Ga(\n +1) \ov \Ga(1+{\l-d \ov 2}+s_1)}
\sum_{m=0}^{k} {(-1)^m k! \ov m! (k-m)!} 
{ \Ga(1+{\l-d \ov 2}+s_1 +m)
\ov ({\te_{34} \ov 2} + m) \Ga(\n+1+m)}
}}
from which we can see that it is convergent and 
has no poles inside the contour in \final. 

Plugging \tra\ into \final, we get, 
$$
S_{\l} = {C \ov \Ga(1+{\te_{34} \ov 2})} \, {1 \ov 2 \pi i} \int_{{\cal C}}
\! d s_1 \, \x^{-s_1} \, \Ga({\l_{34}\ov 2} - s_1)
 \Ga ({\l \ov 2}-s_1) \, H(\l, s_1)  
$$
\eqn\refina{
\times  \, {_3}F_{2} ({\te_{34} \over 2}, 1+{\l-d \ov 2}+s_1,
1-{\ep_{12} \ov 2}; 1+ {\te_{34} \over 2}, \n+1;1)
}
where $C$ is given below eq \final\ and $H(\l,s_1)$ is defined by,
$$
H(\l,s_1) = {\Ga( {\De_{34} \ov 2} + s_1) \Ga({\De_{43} \ov 2} + s_1)
\Ga( {\De_{12} \ov 2} + s_1) \Ga({\De_{21} \ov 2} + s_1) 
\ov  \Ga (2s_1)}
F( {\De_{34} \ov 2} +s_1 ,{\De_{12} \ov 2} + s_1; 2s_1; 1-{\e \ov \x}) \ . 
$$
Naively we may expect  from \final\ that there
are  double poles at $s_1 = {\l_{12} \ov 2} + n, \, n=0,1, \cdots$.
However, eq \refina\ indicates that they are actually  simple poles. This 
result may  also be seen indirectly from eqs \bba\ and 
\aaax\ -- \ggnt: there is no singularity developed in either 
\bba\ or \aaax\ when ${\ep_{12}\ov 2}$ approaches a positive integer. 
In fact it can be checked that the residue 
of the integrand of \refina\ at a pole $s_1 = {\l_{12} \ov 2} + n$ 
is equal to the sum of \bba\ and \aaax\ at the corresponding pole.
If further ${\ep_{34} \ov 2}$ is an integer, then from \refina\ the pole at 
$s_1 ={\l \ov 2} + k+1 + m = {\l_{12} \ov 2} + m = {\l_{34} \ov 2} + n$
($m$ and $n$ non-negative integers) is a double pole instead of a triple pole.
In particular, there is no terms proportional to $(\log \x)^2$ here \freedh.
Similar analysis can be applied when ${\ep_{34} \ov 2}$ is a positive integer.

The appearance of logarithm in coinciding pole cases can be summarized 
as follows:

(i) Only one of ${\ep_{12} \ov 2}$,  ${\ep_{34} \ov 2}$ or
${\l_{12}-\l_{34} \ov 2}$ is an integer:

(i.1) ${\l_{12}-\l_{34} \ov 2}$ is an integer: $\log \x$ associated 
with the double poles at $s_1 = {\l_{34} \ov 2} + n $.

(i.2) ${\ep_{12} \ov 2}$ or  ${\ep_{34} \ov 2}$ is a positive integer:
all poles are simple poles, no logarithm.

(i.3) ${\ep_{12} \ov 2}$ or  ${\ep_{34} \ov 2}$ is zero or a negative integer:
$\log \x$ associated with the double poles at $s_1 = {\l \ov 2} + n $.

(ii) ${\ep_{12} \ov 2}$,  ${\ep_{34} \ov 2}$ and
${\l_{12}-\l_{34} \ov 2}$ are all integers:

(ii.1) at least one of ${\ep_{12} \ov 2}$ and
${\ep_{34} \ov 2}$ is  positive: except for a finite number of simple
poles, all poles are double poles with  $\log \x$.

(ii.2)  ${\ep_{12} \ov 2}$ and ${\ep_{34} \ov 2}$ are zero or negative:
except for a finite number of them, all poles are triple poles with 
$(\log \x)^2$.

\newsec{Scattering amplitudes from generic vertices  and contact terms}

Scattering amplitudes from more complicated interaction vertices such as
$\p \del_\m \p_1 \del^\m \p_2 $ and $\p D_ \n \del_\m \p_1 D_\n 
\del^\m \p_2 $ can be reduced to \fourt\ and contact-type
interactions by integration by part \liut\
or field redefinitions \sei. 
For example, the amplitude resulting from vertices 
$\p \del \p_1 \del \p_2$ and $\p \p_3 \p_4$ can be written as ($d \a$ and 
$d \b$ denote the integration measures as in \fourt),
\eqn\deriv{\eqalign{
& \int \! d \a d \b \,\, \del \KK_1 \, \del \KK_2 \,\,
G(x,y) \,\, \KK_3 \, \KK_4 \cr
& = \ha  \int \! d \a d \b \,\, \big[ \, \del^2 (\KK_1 \,  \KK_2) - \del^2 \KK_1
\, \KK_2 -  \del^2 \KK_2 \, \KK_1 \big ]\,\,  G(x,y) \,\, \KK_3 \, \KK_4 \cr
& = - \ha \int \! d \a d \b \,\, \KK_1 \, \KK_2 \, \KK_3 \, \KK_4 
+ \ha (m^2 - m_1^2 -m_2^2) 
\int \! d \a d \b \,\,  \KK_1 \, \KK_2 \,\, 
G(x,y) \,\, \KK_3 \, \KK_4
}}
Note the coefficient of  the second term  $\ha (m^2 - m_1^2 -m_2^2)$
is precisely the ratio between coefficients of 
$<\Phi_\l \Phi_{\l_1} \Phi_{\l_2}>$  calculated from two types of 
interactions $\p \del \p_1 \del \p_2$ and $\p  \p_1  \p_2$ \free. 

In general we can consider the following Lagrangian of scalar fields,
\eqn\slagr{
{\cal L} = \ha (\del \p_i)^2 + \ha m_i^2 \p_i^2 +
 A^{(0)}_{ijk} \p_i  \p_j  \p_k + 
A^{(1)}_{ijk} \p_i D^{\m} \p_j D_{\m} \p_k
+ \cdots +  A^{(n)}_{ijk} \p_i D^{(n)} \p_j  D^{(n)} \p_k  
}
where $D^{(m)} \p_i$ is defined by
\eqn\genint{
D^{(m)} \p_i = D_{\{\m_1}D_{\m_2} \cdots D_{\m_m \}} \p_i \ .
}
$\{ \,\, \}$ in \genint\  denotes that the indices are symmetrized 
and traces are removed.\foot{We can use $D^2 \p = m^2 \p + \cdots$ 
to reduce terms containing traces of indices to  lower order terms.
Similarly the commutators of derivatives $[D_\m, D_\n] \propto R$ 
also reduce to lower order terms, where $R$ is the constant curvature.}   
For the purpose of tree-level four-particle scattering we can 
eliminate those vertices with derivatives  
by a field redefinition, 
\eqn\fiere{
\p_i = \p'_i + B^{(0)}_{ijk} \p'_j \p'_k + 
\cdots +  B^{(n-1)}_{ijk}  D^{(n-1)} \p'_j \,  D^{(n-1)} \p'_k  \ .
}
$B$'s in \fiere\ can found by plugging \fiere\ into \slagr\ and setting 
to zero the coefficients of the cubic derivative vertices.
The resulting  Lagrangian can be written as,
\eqn\lagar{\eqalign{
{\cal L} & = \ha (\del \p'_i)^2 + \ha m_i^2 (\p'_i)^2 + 
\l_{ijk} \p'_i  \p'_j \p'_k \cr
& + {\rm contact \,\, vertices
\, \, of \,\, quartic \,\,or \,\, higher \,\, order}
}}
For example for $n=2$ in AdS$_{d+1}$, $B$'s can be found to be \sei,
$$
B^{(1)}_{ijk} = \ha A^{(2)}_{ijk}, \,\,\,\;\;\;\;\;\;
B^{(0)}_{ijk} = \ha A^{(1)}_{ijk} + \four  A^{(2)}_{ijk} (m_i^2 -
m_j^2 -m_k^2 + 2d) 
$$
and 
\eqn\pronm{
\l_{ijk} =  A^{(0)}_{ijk} + B^{(0)}_{ijk}(m_i^2 -
m_j^2 -m_k^2) - {2 \ov d+1} m_j^2 m_k^2 B^{(1)}_{ijk} \  .
}

Thus for  generic interactions, the scattering amplitude 
can be written as, 
\eqn\split{
A_{\l} = S_{\l}  + S^{(4)}
}
where $S_{\l}$ is given by \final\ with normalised vertices 
\pronm\ and  $S^{(4)}$ is given by quartic vertices 
in \lagar.

Let us now look at the contribution from contact terms. 
We observe that by repeatedly using the identity 
($J_{\m \n}(x) = \d_{\m \n} - 2 x^\m x^\n / |x|^2 $),
\eqn\contai{\eqalign{
& D^{\m} \KK_{\l_i} (u,x_i) D_{\m} \KK_{\l_j} (u,x_j) =
c_{\l_i} c_{\l_j} \,u_0^2 \del_{\m}  ({u_0 \over |u-x_i|^2})^{\l_i} \,
\del_{\m} ({u_0 \over |u-x_j|^2})^{\l_j} \cr
& = \l_i \l_j \, \, \KK_{\l_i} (u,x_i)\KK_{\l_j} (u,x_j)
J_{\m 0} (u-x_i) J_{\m 0} (u-x_j) \cr
& = \l_i \l_j \,  \KK_{\l_i} (u,x_i)\KK_{\l_j} (u,x_j)
- 2 \n_i \n_j \, x_{ij}^2 \, \KK_{\l_i+1} (u,x_i)\KK_{\l_j+1} (u,x_j)
}}
and $D^2  \KK_{\l_i} = m_i^2  \KK_{\l_i}$
we can put a generic quartic contribution into a sum of 
terms without derivatives,
\eqn\da{\eqalign{
S_{c} & = \int \! {d u_0 d^d u \over u_0^{d+1}} \
\KK_{\l_1} (u,x_1) \KK_{\l_2} (u,x_2)  
\KK_{\l_3} (u,x_3) \KK_{\l_4} (u,x_4)  \cr 
& = \Pi_i c_{\l_i} \, 
\int \! {d u_0 d^d u \over u_0^{d+1}} \
({u_0 \over |u-x_1|^2})^{\l_1} ({u_0 \over |u-x_2|^2})^{\l_2}
({u_0 \over |u-x_3|^2})^{\l_3}
({u_0 \over |u-x_4|^2})^{\l_4}. 
}}
Thus it is enough to look at \da.

Contact contribution \da\ in \ads\  have been discussed 
before in \muck\ and \refs{\freedm} (see also \refs{\brodie}). In 
particular in \freedm\ it was pointed out that when $\l_{12} =\l_{34}$
the leading term in short distance limit $x_{12}, x_{34} \ra 0$
is given by a logarithmic contribution. 
Here we give a more thorough analysis of the analytic properties
of \da, presenting the result in a way suitable for our later
discussion of its \cft\ interpretation.

In Appendix C, we show that similarly to the exchange amplitude, the 
contact contribution \da\ can also be written as an inverse 
Mellin integral, 
$$
S_c  = C_c \, {1 \ov 2 \pi i} 
\int_{\C}
\! d s \, \x^{-s} \,\,
 \Ga({\l_{12} \ov 2} -s) 
\Ga({\l_{34}\ov 2} - s) \,
F( {\De_{34} \ov 2} +s ,{\De_{12} \ov 2} + s; 2s; 1-{\e \ov \x}) 
\
$$
\eqn\contacf{
\times \,\,
{\Ga( {\De_{34} \ov 2} + s) \Ga({\De_{43} \ov 2} + s)
\Ga( {\De_{12} \ov 2} + s) \Ga({\De_{21} \ov 2} + s) 
\ov  \Ga (2s)}
}
with 
$$
C_c = {1 \ov 2 \pi^{3d/2}} {\Ga({\l_{12} + \l_{34} -d \ov 2})
\ov \Ga(\n_1)\Ga(\n_2)\Ga(\n_3)\Ga(\n_4)} 
{1 \ov |x_{12}|^{\l_{12}} \, |x_{14}|^{\De_{12}} \,
|x_{24}|^{\De_{21} - \De_{34}} \, |x_{23}|^{\De_{34}} \, 
|x_{34}|^{ \l_{34}}}
$$
where the integration path $\C$ should be understood in the same sense
as that in \jans\ (see the remark below \jans). 
Thus in the $s$-channel limit $\e, \x \gg 1$, \contacf\ can be written 
as, 
$$
S_c = \sum_{n=0}^{\infty} S_{cn}^{(\l_{12})} + \sum_{n=0}^{\infty} S_{cn}^{(\l_{34})} 
$$
where, 
$$
S_{cn}^{(\l_{12})} = 4 A^{(\l_{12})}  {(-1)^{n} \ov n!} 
\x^{-n} \,\,
F({\l_{12} + \De_{34} \ov 2}+ n, \l_1 + n; \l_{12} + 2n; 1-{\e \ov \x})\,
\times
$$
\eqn\conaa{
\Ga ({\l_{34}-\l_{12} \ov 2} -n)\,
{\Ga ({\l_{12} + \De_{34} \ov 2}+ n) 
\Ga ({\l_{12} - \De_{34} \ov 2}+ n)
\Ga (\l_1 + n) \Ga (\l_2 + n)  \ov \Ga(\l_{12} + 2n)}
}
and $ S_{cn}^{(\l_{34})}$ can be obtained from \conaa\ by taking 
$1,2 \ra 3,4$. Note $A^{(\l_{12})}$ in above is given by eq \coefa\ 
and  except for the extra $G_n$ in \aaax, 
\conaa\ is almost identical to \aaax.

When ${\l_{12} - \l_{34} \ov 2}$ is an integer,  
except for a  finite number of poles the two ascending 
simple-pole sequences of 
the integrand in \contacf\ will merge into a 
double-pole sequence. 
Again as in the case of exchange amplitude, the double-pole contribution
will contain $\log \x$. 
In particular, when $\l_{12} = \l_{34}$
all ascending poles become double poles and the leading contribution 
contains a $\log \x$. 

Note that since \da\ is symmetric under exchanges of its four boundary  
propagators, its expansion in $u-$channel limit $x_{13}, x_{24} \ra 0$
can be obtained by exchanging $2$ and $3$ in \contacf\ and \conaa\ and  
$\x \ra {\x \ov \e}$ and $\e \ra {1 \ov \e}$.

\newsec{Four-point functions and conformal partial wave expansion 
in CFT}

To seek a \cft\ interpretation of the \ads\ amplitudes 
discussed in last two sections, in this section we 
review the  conformal partial
wave expansion (CPWE) approach to the calculation of 
four-point functions in CFT \refs{\ferr,\cpw} (for a review 
see \refs{\fpf,\tmp}, see also \refs{\petkou} for some recent 
discussions\foot{I would like to thank A. Petkou for bringing these 
references to my attention.}). 

In CFT$_d$, the states generated by acting  by a  product of 
the conformal operators on the vacuum can be decomposed into a 
direct sum of irreducible representations of the conformal group
\eqn\wave{
\P_{1}(x_1) \P_{2}(x_2)|0> = 
\sum_k \int d^d x \, \, Q_{12k}(x|x_1,x_2)
|k, x>\ ,
}
where $k$ sums over all the irreducible representations in the Hilbert space 
and states $|k, x> = \P_k (x) |0>$ span the space of an irreducible   
representation of the conformal group. 
\wave\ can be further lifted into an operator equation,
\eqn\ope{
\P_{1}(x_1) \P_{2} (x_2) = \sum_k \int d^d x \, \, 
Q_{12k}(x|x_1,x_2)
\Phi_{k} (x)\ , 
}
understood as a relation between correlation functions.  
The summation in \ope\ is over primary fields (non-derivatives) only and 
the integration over all space effectively incorporated the contribution 
of their  $SO(d,2)$ descendants (fields with derivatives).  
The short-distance OPE  
can be obtained from \ope\ in small $|x_{12}|$ limit by
expanding the integrand  in terms of $x_1-x_2$.
When $\P$'s are orthogonal to each other, it can be seen from \wave\
that $Q$'s are given by the amputated three-point functions.

Applying \wave\ to a  four-point function we find 
$$
W_{1234}(x_1,x_2,x_3,x_4) = \ 
<0|\P_{1}(x_1)\P_{2}(x_2)\P_{3}(x_3) \P_{4}(x_4)|0>
$$
\eqn\parf{
= \sum_k \int \! d^d x d^d y \, \, Q_{12k}(x_1,x_2|x)\ 
W_{k}(x-y)\  Q_{k34}(y|x_3,x_4)\ . 
}
where $W_k (x-y) = <0|\P_k (x) \P_k (y)|0>$. 

In the following, we shall look at the contribution of 
an intermediate scalar operator $\P_{\l}$ (with dimension $\l$)
to the four-point function of four scalar operators 
$\P_{\l_i},  i=1, \cdots, 4$ (with  dimensions $\l_i$ respectively),
\eqn\sinco{
S_{\l} = \int \! d^d x d^d y \, \, Q_{\l \l_1 \l_2}(x_1,x_2|x)\ 
W_{\l}(x-y)\  Q_{\l \l_3 \l_4}(y|x_3,x_4)\ .
}

In Euclidean signature, the two-
and three-point functions are given by,
$$
G_\l (x-y) = <\P_\l (x) \P_\l (y) > = {c \ov |x-y|^{2 \l}}, 
\,\,\,
$$
$$
G_{\l \l_1 \l_2} (x, x_1, x_2) = 
< \P_{\l} (x) \P_{\l_1} (x_1) \P_{\l_2}(x_2) > = 
f_{\l \l_1 \l_2} \, A_{\l \l_1 \l_2} (x,x_1,x_2), \,\,\,
$$
$$
G_{\l \l_3 \l_4} (x, x_3, x_4) = 
< \P_{\l} (x) \P_{\l_3} (x_3) \P_{\l_4}(x_4) > =
f_{\l \l_3 \l_4}  A_{\l \l_3 \l_4} (x,x_3,x_4) 
$$
where function $A_{abc} (x,y,z)$ is defined by,
$$
A_{abc} (x,y,z) = {1 \ov |x-y|^{a+b-c} |z-y|^{c+b-a}|x-z|^{a+c-b}}.
$$
The normalisation constants $c$ and $f$'s 
will be taken to be those given by AdS calculations \free \foot{Here the 
normalisation for three-point functions is given
by interaction vertex $\p_1 \p_2 \p_3$ in \ads. 
When considering more complicated vertices,
an additional normalisation factor \pronm\ should be taken into 
account.}, i.e.
\eqn\ttnor{
c = {\Ga(\l) \ov \pi^{d/2} \Ga(\n)} (2 \l -d), \,\,\,
 \;\;\;\;\;\;
f_{\l \l_1 \l_2} =  { \Ga({\ep_{12} \ov 2}) \Ga({\te_{12} \ov 2})
\Ga({\d_1 \ov 2}) \Ga({\d_2 \ov 2}) \ov 2 \pi^{d} \Ga(\n_1)
\Ga(\n_2) \Ga(\n)}
}
and  a similar expression for $f_{\l \l_3 \l_4}$ obtained 
from $f_{\l \l_1 \l_2}$ by taking $1,2 \ra 3,4$.

In Minkowski signature, due to the spectrality condition, it is more 
convenient to work in momentum space, where the two- and three-point functions 
are given by
\eqn\zzz{
W(p) = -i {\rm Disc} \, G(p)|_{p^{d} = -i p^0}
={ 2 \pi \ov \Ga(1+\n) \Ga(-\n)} \t(p^0) \t(-p^2_{Min}) 
G(p)|_{p^{d} = -i p^0}
}
\eqn\zzy{
W(p|x_1, x_2) = -i {\rm Disc} \, (p|x_1,x_2)|_{p^{d} = -i p^0}
}
where $G(p)$ and $G(p|x_1,x_2)$ are Euclidean two- and three-point functions 
in momentum space,
$$
G(p) = c \int \! d^d y \, e^{-i p \cdot y} {1 \ov |y|^{2 \l}}
= c {\pi^{{d \ov 2}} \Ga(- \n) \ov 2^{2 \n} \Ga (\l) } p^{2 \n}
$$
$$
G_{\l \l_1 \l_2}(p|x_1,x_2) = \int \! d^d y \, e^{-i p \cdot y} 
G_{\l \l_1 \l_2}(y,x_1,x_2) =
f_{\l \l_1 \l_2}
 {2 \pi^{{d \ov 2}} \ov \Ga ({\d_{1} \ov 2}) 
\Ga ({\d_{2} \ov 2})} {1 \ov x_{12}^{\ep_{12}}} 
({p^2 \ov 4 x_{12}})^{{\n \ov 2}} \, \times
$$
\eqn\eucth{
\int_{0}^{1} \! du \, u^{{\De_{12} \ov 2} + {d \ov 4} -1}\,
(1-u)^{{\De_{21} \ov 2} + {d \ov 4} -1} \,
e^{-i p \cdot [ux_1 +(1-u) x_2]} \, K_{\n} (\sqrt{u(1-u) p^2 x_{12}^2}\,)
}
The amputated three-point function $Q$ can then be found to be,
$$
Q_{\l \l_1 \l_2} (p|x_1,x_2)  = W_{\l}^{-1}(p) W_{\l \l_1 \l_2}(p|x_1, x_2) 
= c^{-1}  f_{\l \l_1 \l_2} {2^{\n} \Ga(\l) \Ga(1+ \n) \ov 
\Ga ({\d_{1} \ov 2}) \Ga ({\d_{2} \ov 2})} 
{p^{-\n} \ov x_{12}^{\l_{12}-{d \ov 2}}}  \times
$$
\eqn\zzx{
\int_{0}^{1} \! du \, u^{{\De_{12} \ov 2} + {d \ov 4} -1}\,
(1-u)^{{\De_{21} \ov 2} + {d \ov 4} -1} \,
e^{-i p \cdot [ux_1 +(1-u) x_2]} \, I_{\n} (\sqrt{u(1-u) p^2 x_{12}^2}\,)
}
In \eucth\ and \zzx\ $K_{\n}$ and $I_{\n}$ are modified Bessel functions.

Plugging eqs \zzz\ and \zzx\ into \sinco, we have, in momentum space
with Minkowskian signature,
\eqn\zzw{
W_{\l} = {1 \ov 2 \pi^{d}} \int \! d^d p \,
Q^*_{\l \l_1 \l_2} (p|x_1,x_2) \, W_{\l} (p) \, Q_{\l \l_3 \l_4} (p|x_3,x_4) 
}
The integrals in \zzw\ were explicitly computed in \ferr\ and the result 
can written as an inverse Mellin integral,
$$
W_{\l} = c^{-1} f_{\l \l_1 \l_2} f_{\l \l_3 \l_4}
{ \Ga (\l) \Ga(1 + \n) \ov \Ga ({\d_{1} \ov 2}) \Ga ({\d_{2} \ov 2})
\Ga ({\d_{3} \ov 2}) \Ga ({\d_{4} \ov 2})}
{1 \ov  |x_{12}|^{\ep_{12}} \, |x_{14}|^{ \d_1} \,
|x_{24}|^{\De_{21} - \De_{34}} \, |x_{23}|^{\d_3} \, 
|x_{34}|^{ \ep_{34}}} \times
$$
\eqn\mincpw{{1 \ov 2 \pi i} \int_{-i \infty}^{i \infty} \! 
ds \, (-\x)^{-s}\, \Ga (-s) {\Ga ({\d_1 \ov 2}+ s) \Ga ({\d_2 \ov 2}+ s)
\Ga ({\d_3 \ov 2}+ s) \Ga ({\d_4 \ov 2}+ s)
\ov \Ga(\l + 2s) \Ga (\n +s +1)}
F({\d_3 \ov 2}+ s, {\d_1 \ov 2} + s; \l + 2s; 1-{\e \ov \x})
}
where $\x$ and $\e$ are cross ratios defined in \crosr\
 and the Mellin integral 
should be understood in the same sense as the ones in previous 
sections.
Again when $\e,\x >1$, \mincpw\ can be written as an expansion,
$$
W_\l = {1 \ov 8 \pi^{{3 \ov 2} d}} 
{\Ga({\ep_{12} \ov 2})\Ga({\ep_{34} \ov 2})
\Ga({\te_{12} \ov 2})\Ga({\te_{34} \ov 2})
\ov  \Ga(\n_1) \Ga(\n_2)  \Ga(\n_3)\ \Ga(\n_4)}
{1 \ov  |x_{12}|^{\ep_{12}} \, |x_{14}|^{ \d_1} \,
|x_{24}|^{\De_{21} - \De_{34}} \, |x_{23}|^{\d_3} \, 
|x_{34}|^{ \ep_{34}}} \times
$$
\eqn\minfcp{
\sum_{n=0}^{\infty}  {1 \ov n! } 
\x^{-n} {\Ga ({\d_1 \ov 2}+ n) \Ga ({\d_2 \ov 2}+ n)
\Ga ({\d_3 \ov 2}+ n) \Ga ({\d_4 \ov 2}+ n)
\ov \Ga(\l + 2n) \Ga (\n +n +1)}
F({\d_3 \ov 2}+ n, {\d_1 \ov 2} + n; \l + 2n; 1-{\e \ov \x}) 
}
We notice that \minfcp\ agree precisely with \bba\
including the numerical coefficient.

\newsec{\cft\ interpretation of \ads\ amplitudes }

In previous sections, we have managed to express all our results 
as inverse Mellin integrals and when 
$\x, \e >1$ 
write them  in terms of inverse power series
of $\x,\e$ as a sum of residues of the integrand. 
When the pole sequences in \polese\ and \contacf\ do not coincide 
with one another, in all cases (see eqs \bba,\aaax,\conaa) 
the contribution from a pole sequence
can be written in a similar pattern as the CPWE expression \minfcp,
\eqn\sumres{
\sum_{n=0}^{\infty}  a_n \,  H(\L+2n)
}
where each term in the summation is given by the residue at the pole 
${\L \ov 2} + n$. In \sumres, $a_n$ are  numerical  
coefficients and $H$ is a function defined by,
$$
H(\a) = {1 \ov  |x_{12}|^{\l_{12}-\a} \, |x_{14}|^{\a + \De_{12}} \,
|x_{24}|^{\De_{21} - \De_{34}} \, |x_{23}|^{\a + \De_{34}} \, 
|x_{34}|^{\l_{34}-\a}} \times
$$
\eqn\patter{
{\Ga ({\De_{12} \ov 2}+ {\a \ov 2}) \Ga ({\De_{21} \ov 2}+ {\a \ov 2})
\Ga ({\De_{34} \ov 2}+ {\a \ov 2}) \Ga ({\De_{43} \ov 2}+ {\a \ov 2})
\ov \Ga(\a)}
F({\De_{34} \ov 2}+ {\a \ov 2}, 
{\De_{12} \ov 2} + {\a \ov 2}; \a; 1-{\e \ov \x}) 
} 
In \minfcp, 
\eqn\cpweco{
a_n = {1 \ov 8 \pi^{{3 \ov 2} d}} 
{\Ga({\l_{12} - \L \ov 2})\Ga({\l_{34} -\L \ov 2})
\Ga({\l_{12} + \L -d \ov 2})\Ga({\l_{34} + \L -d \ov 2})
\ov \Ga(\n_1)\Ga(\n_2)\Ga(\n_3)\Ga(\n_4)}  
{1 \ov \Ga(n+1) 
\Ga (\L -{d \ov 2} + n +1)}.
}
The contact amplitude  \da, may be written as,
\eqn\condec{
S_c = S_c^{(\l_{12})} + S_c^{(\l_{34})}
}
where  $S_c^{(\l_{12})}$, $S_c^{(\l_{34})}$ are of the form of \sumres\ 
with $\L = \l_{12}, \l_{34}$. For  $S_c^{(\l_{12})}$, 
\eqn\coefi{
a_n = {1 \ov 2 \pi^{3d/2}} {\Ga({\l_{12} + \l_{34} -d \ov 2})
\ov \Ga(\n_1)\Ga(\n_2)\Ga(\n_3)\Ga(\n_4)}   
{(-1)^n  \ov \Ga(n+1) } \Ga ({\l_{34} - \l_{12} \ov 2} -n)
}
with those of $S_c^{(\l_{34})}$ given by \coefi\ 
with $\l_{12} \lra \l_{34}$.
Similarly, the exchange amplitude \fourt\ 
can be written in terms of \sumres\ with 
$\L = \l, \, \l_{12}, \, \l_{34}$,
\eqn\exdim{
S_{ex} = S^{(\l)} + S^{(\l_{12})} + S^{(\l_{34})}
}
where $a_n^{(\l)}$ are the same as those of CPWE  \cpweco\ and 
\eqn\coefii{
a_n^{(\l_{12})} ={1 \ov 8 \pi^{3d/2}} {\Ga({\l_{12} + \l_{34} -d \ov 2})
\ov \Ga(\n_1)\Ga(\n_2)\Ga(\n_3)\Ga(\n_4)}
  {(-1)^n  \ov \Ga(n+1) } \Ga ({\l_{34} - \l_{12} \ov 2} -n) G_n
}
with $G_n$ given by \ggn\ or \ggnt. 
$a_n^{(\l_{34})}$ may be obtained from 
\coefii\ with $1,2$ and $3,4$ exchanged.

To have a more precise picture of what we have found so far,
we would like to understand the physical 
meaning of the poles and the residues of each pole.
For this purpose, let us go back to the contribution of a primary
operator $\P_\L$ to the operator algebra \ope, 
\eqn\waop{
\P_{\l_1}(x_1) \P_{\l_2} (x_2) =  \int d^d x \, \, 
Q_{\L \l_1 \l_2 }(x|x_1,x_2)
\Phi_{\L} (x)\ , 
}
where 
\eqn\froiu{
Q_{\L \l_1 \l_2}(x|x_1,x_2) = \int \! d^d p \, e^{i p \cdot x} 
Q_{\L \l_1 \l_2}(p|x_1,x_2)
}
and $Q_{\L \l_1 \l_2}(p|x_1,x_2)$ is given by \zzx. 
For simplicity we will look at the 
analytic continuation of \zzx\ to Euclidean space\foot{This is not the 
completely right thing to do,  as in Minkowski signature integration 
over momenta involves the spectrality conditions
$p_0 >0, p^2 <0$, in which case the  expressions are  rather complicated.
For illustrative purpose, we shall use an Euclidean expression.}.
Plugging \zzx\ and \froiu\ into \waop, we find that \refs{\ferrg},
\eqn\wwwi{\eqalign{
\P_{\l_1}(x_1) \P_{\l_2} (x_2) & = \sum_{n=0}^{\infty}
b_n  {1 \ov x_{12}^{\l_{12} - \L - 2n}}
\int_{0}^{1} \! du \, u^{{\d_1 \ov 2} + n-1}  \, (1- u)^{{\d_2 \ov 2} + n-1} \,
e^{u x_{12} \cdot \del} \,
\P_{\L}^{(n)}(x_2) \cr
& = \sum_{n=0}^{\infty} b_n F_n
 } }
where $b_n$ are numerical factors and we have defined operators,
\eqn\subpri{
\P_{\L}^{(n)}(x_2) = (\del^2)^n \P_\L (x_2), \,\,\,\,\,\, n=0,1,2,\cdots
}
which  we will call sub-primary operators (n=0 is primary).
To reach \wwwi, we have used the series 
expansion for Bessel function.
$F_n$ in \wwwi\ denotes the contribution to OPE from  a sub-primary
operator  $\P_{\L}^{(n)}$ and its diagonal descendants. 
By diagonal descendants we mean the states in weight diagram 
diagonally generated from a sub-primary (see figure 5).
We may now interpret that each pole in \mincpw\ represents 
the dimension of a sub-primary and the residue at the pole corresponds 
to the contribution of the sub-primary and its diagonal descendants.

\fig{
Weight diagram for $SO(d,2)$ representation $D(\l, 0)$. The horizontal
axis $l$ represents the set of indices for $SO(d)$ subgroup, and
the perpendicular axis represents  the scale dimension. A sub-primary
is a state lying on the line with $l=0$. The states connected to a 
sub-primary by dotted lines are descendants diagonally generated from
the sub-primary.
}{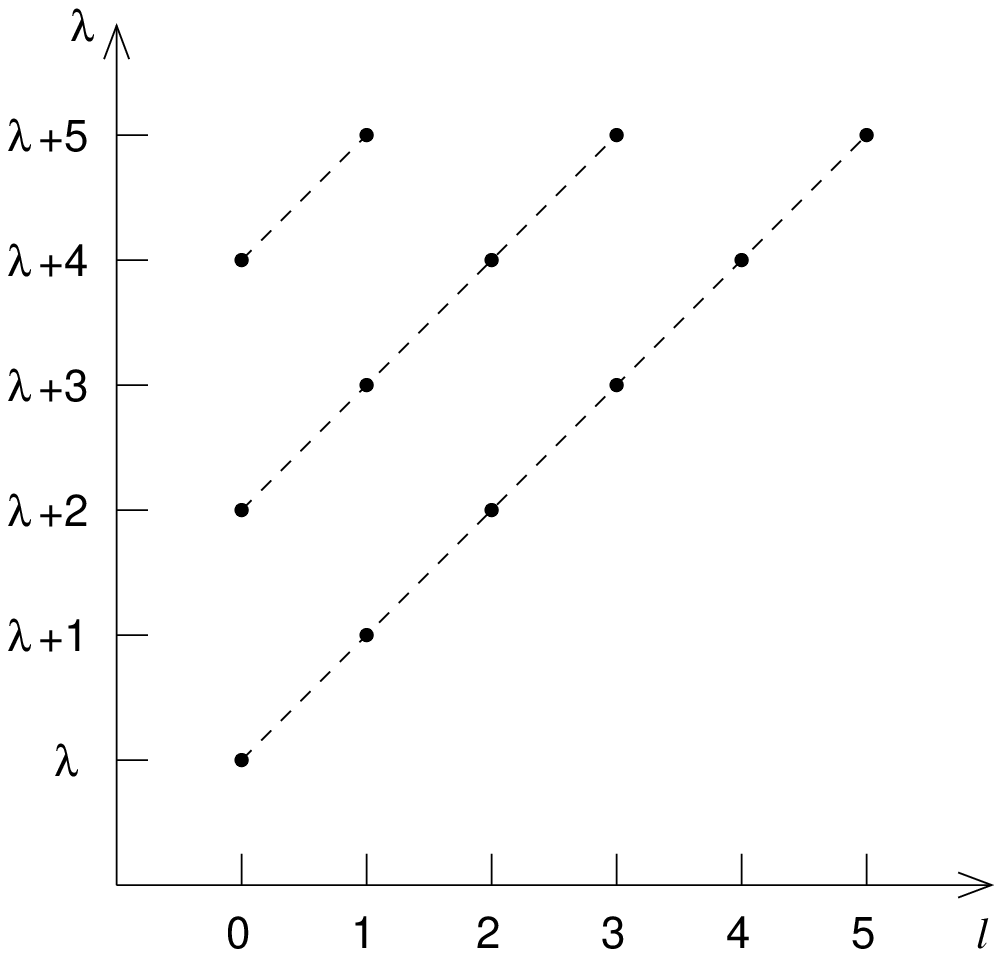}{5.8 truecm}
\figlabel\weight

By comparing \exdim\ and \condec\ with CPWE \sumres\ and \cpweco, 
we see that for the case of non-coinciding poles:

i) $S^{(\l)}$ in exchange amplitude \exdim, which arises from the first 
pole sequence ${\l \ov 2} + n$
in \polese\ agrees  precisely with the result from conformal partial wave 
expansion \minfcp, including the overall numerical coefficient.
This indicates that $S^{(\l)}$  has a \cft\ interpretation  in terms of 
exchange of a primary operator of dimension $\l$.

ii) In both exchange
and contact amplitudes, the contributions from a pole 
at ${\l_{12} \ov 2} + n $ (or ${\l_{34} \ov 2} + n $)
agree exactly with that of a sub-primary operator 
with dimension $\l_{12} + 2n$ ( $\l_{34} + 2n$) and its 
diagonal descendants. 
But the numerical coefficients \coefii\ and \coefi\ at each  pole are
different from that predicted by CPWE \cpweco, in particular, in 
exchange amplitude  the coefficient involves a somewhat complicated 
factor $G_n$ \ggn.

The above identification  of \condec\ and \exdim\ in terms of exchanges 
of  sub-primary operators also cast light on the 
conditions \logex\ and \logcon\ for the occurrence of logarithms, which
are satisfied precisely when sub-primaries of  different operators 
become degenerate. 
For example, consider a contact diagram with 
${\l_{12} - \l_{34} \ov 2} =k$ and $k$ a positive integer. The dimension
$\l_{12} + 2n$ of a  sub-primary operator $O^{(n)}_{12}$ of 
$O_{12}$ will then be the same as that of the sub-primary $O^{(k+n)}_{34}$. 
In addition,  the quantum numbers of all the diagonal descendants 
generated from $O^{(n)}_{12}$ and $O^{(k+n)}_{34}$ will be identical (see
figure \weight). 
To see this more explicitly, let us move slightly off the 
degeneracy point, i.e. consider, $\l_{12} = \l_{34} + 2k + 2 \ep$ where  
$ 0 < \ep < 1$.  
Then \coefi\ may be written as (below we will omit the overall constant),
$$
a_n^{12} = (-1)^k {\pi \ov \sin \ep \pi}{1 \ov \Ga(n+1) \Ga (1+n+k+\ep) } 
$$
where we have used $\Ga(x) \Ga(1-x) = {\pi \ov \sin x \pi}$.
And \condec\ may be written as,
$$
S_c = \sum_{n=0}^{k-1} \, {(-1)^n \ov n!} \, \Ga(k-n+\ep) \, 
H(\l_{34} + 2n) \,\, 
$$
\eqn\sysdif{ 
+ \,\, 
(-1)^k {\pi \ov \sin \ep \pi} \sum_{m=0}^{\infty} 
[ {H(\l_{34}+ 2m+2 k + 2 \ep) \ov \Ga(m+1 ) \Ga (1+m+k+\ep) } \, -\,
{H(\l_{34}+ 2m+2k) \ov \Ga(m+1-\ep) \Ga (1+m+k) }  ]
}
As we take $\ep$ to zero, the conflicting  
contributions from $O_{12}^{(m)}$ and  $O_{34}^{(m+k)}$ in the square 
bracket become degenerate and the result is 
given by their derivatives over $\ep$, which contain logarithms.
The above discussion suggests that  by turning on a very small $\ep$
at degeneracy points, \sysdif\ provides a useful  
way to ``regularize'' logarithms. 

Since a contact diagram is symmetric under exchanging
its external legs, it's  $t-$ or  $u-$channel expansion  
can be simply obtained by taking $2 \lra 4$ or $2 \lra 3$ in \condec\ and
\coefi. For example, it can be represented as $t-$channel 
exchange in \cft\ as  in figure 6. 

\fig{
$t-$channel OPE interpretation  of a contact diagram. 
}{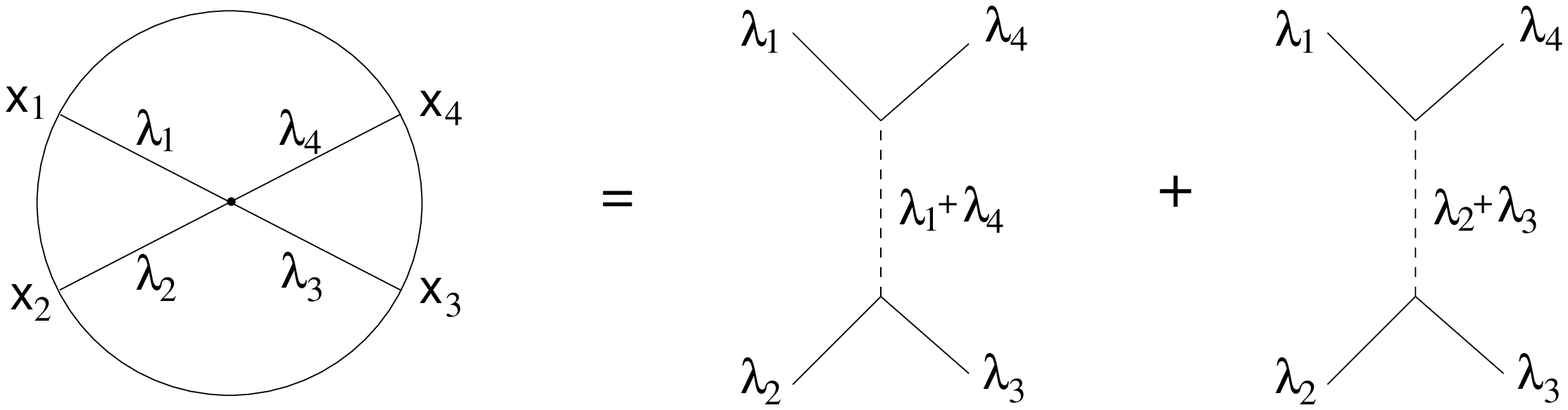}{11.8 truecm}
\figlabel\cfticu

\medskip\noindent
{\it Discussions}
\medskip

We note that conformal symmetry imposes strong restrictions
on the coefficients $C^k_{ij}$ in \wope; that of primary determines 
those of  their descendants. The structure of \wwwi\ and \minfcp, 
including the numerical coefficients $a_n$ in \cpweco\ and $b_n$ in \wwwi\ 
in the summation,  are uniquely 
fixed up to an overall constant by the fact that 
$\P_{\L}$ and its derivatives
fill an irreducible representation of the conformal group. 
Although 
we have found in \condec\ and \exdim\ the contributions from 
a complete set of sub-primary  operators of dimensions $\l_{12}$ and 
$\l_{34}$, their relative OPE coefficients \wope\ are not 
consistent with those required by conformal symmetry, in other words,
these sub-primaries  do not seem to fill the same irreducible 
multiplets.

It is probably not surprising that  we do not find  a complete 
\cft\ identification in \condec\ and \exdim. After all, we are only 
looking at a generic diagram in \ads, which hardly makes too much 
sense before we specify a particular theory and add up all the 
diagrams contributing  to a realistic amplitude.  The encouraging
message seems to be that we are indeed able to find a relation between 
an arbitrary scattering diagram and OPE, which indicates 
some kind of universality between a theory in \ads\ and \cft. 

It is  not clear at the present time how much we see here will survive in the 
final expression of a realistic amplitude, in particular, whether 
operators $O_{12}$ and $O_{34}$ will have a consistent \cft\ interpretation
when we add up all the diagrams. 
Let us now consider what could be these operators 
if their contribution do survive in the final expression. 
A clue comes from the consideration of free theory,
where the operator product expansion takes the form 
(see e.g. \refs{\frances}),  
\eqn\freeop{
A(z) B(w) = \overbrace{A(z) B(w)} + :A(z) B(w):
}
where the first term on the right hand side denotes contraction and 
$:A(z) B(w):$ stands for normal-ordered operator whose
explicit form can be obtained from a Taylor expansion,
$$
:A(z) B(w): \, = \sum_{k=0}^{\infty} {(z-w)^k \ov k !} (\del^k A B)(w)
$$ 
In free theory the conformal dimension for $:A(z) B(w):$ is just 
$\l_A + \l_B$. Thus naturally we may expect that $O_{12}$ and $O_{34}$
should be the counter-parts of $:O_{\l_1}O_{\l_2}:$ and $:O_{\l_3} O_{\l_4}:$
in interacting theory. In $\N=4$ Super-Yang-Mills with gauge group 
$SU(N)$, $O_{12}$ and $O_{34}$ may be interpreted as
double-trace operators, i.e. operators of type  $Tr F^2 TrF^2 (x)$.  
Since we do not see a continuous spectrum 
of dimensions in \exdim\ and \condec, we would expect $O_{12}$ 
and $O_{34}$ to correspond in \ads\ to two-particle  bound 
states of supergravity/string theory. This is consistent with the 
expectation \refs{\cole} that, to lowest order  in $1/N$, 
there cannot be any two-particle cut in Yang-Mills four-point functions. 

Finally, we note that since  \coefi\ and \coefii\ involve dimensions of  
other operators (e.g. $\l$ and $\l_{34}$), it suggests a possibility that 
the mismatch in our OPE identification 
may be due to certain underlying mixing and
interaction  between different operators at sub-primary
level. Moreover, when $\ep \ra 0$, the pattern 
indicated in \sysdif\ for the degeneracy of sub-primaries strongly 
reminds us of the behaviour of a two-level system. 
Similarly, by examining the exchange amplitude \exdim\ and \coefi,
we also find that the  pattern near degeneracy points 
is rather like a three-state system.

\appendix{A} {Mellin transformation and analytic continuation}

Here we give an brief introduction to the Mellin transformation 
and how to use it to evaluate integrals.\foot{I would like to thank
T.W.B. Kibble,  F.G. Leppington, and A.B. Zamolodchikov for discussions
related to the content of  this appendix.}

The Mellin transform of a function $g(x)$ is
\eqn\mma{
h(s) = \int_0^{\infty} \! dx \, g(x) \, x^{s-1}
}
If the set of convergence of the integral \mma\ has a nonempty
interior $\a < Re(s) < \b$ and $h(s)$ is analytic in this strip,
we can have the inverse Mellin transformation,
\eqn\mmb{
g(x) = {1 \ov 2 \pi i} \int_{c-i \infty}^{c+ i \infty} \! h(s) x^{-s} ds
}
for all $c$ such that $\a < c  < \b$.

Some well-known integral representations of higher transcendental
functions can be interpreted as (inverse) Mellin transforms,  
for example, 
$$
\Ga(s) = \int_0^{\infty} \! dx \, x^{s-1} \, e^{-x}, \,\,\,\,\,
\Ga(s) \zeta(s) = \int_0^{\infty} \! dx \, x^{s-1} \, (e^{x}-1)^{-1} \ .
$$
and the Mellin-Barnes representation of a  
hypergeometric function,
\eqn\mmd{
F(a,b;c;z) = {\Ga(c) \ov \Ga(a) \Ga(b)} 
 {1 \ov 2 \pi i } \int_{-i \infty}^{i \infty} \! ds \,\, 
{\Ga(a+s) \Ga(b+s) \ov \Ga(c+s)} \Ga(-s) (-z)^s \ .
}
Taking $b=c$ in \mmd\ we get, 
\eqn\dmmd{
F(a,b;b;z) = (1-z)^{-a} = {1 \ov \Ga(a)} 
 {1 \ov 2 \pi i } \int_{-i \infty}^{i \infty} \! ds \,
\Ga(a+s) \Ga(-s) (-z)^s \ . 
}

Generally, to evaluate an integral:
\eqn\mme{
I = \int \! dx \, g(x) \, f(x) 
}
we can  first plug into \mme\ the inverse Mellin transform 
\mmb\ of  $g(x)$, then do the $x$-integral and finally  inverse-Mellin
transform back,
\eqn\mmf{
I =  \int_{c-i \infty}^{c+ i \infty} \! ds \, h(s) \, J(s),
\,\,\,\,\, 
J(s) =  \int \! dx \,  x^s \,  f(x).
}

Normally Mellin transform \mma, \mmb\ is not as convenient 
as Fourier or Laplace transform 
as it requires the functions to be transformed have 
reasonable ``good behaviours'' both at zero
and infinity to ensure the existence of the strip where the transform 
can be defined. 
But for those functions $g$ in \mma\ which have a convenient power series 
expansion (such as hypergeometric functions) Mellin representation is more 
powerful 
since the indices and the coefficients of the expansion are represented by 
the poles 
and the corresponding residues of $h(s)$ in the complex $s$-plane.

Let us  look at a simple example,
\eqn\exam{
I = \int_0^{\infty} \! dx \, \, x^{\n -1} (1 + x)^{-\m} (x+t)^{-\r}
}
The integral is  defined when $Re(\n) >0$, $Re(\m + \r -\n) >0$, 
and  $\arg(t) < \pi$ which we assume is the case.
For convenience we will also take $|t|<1$.
The result of \exam\ is well known, given by 
a hypergeometric function, 
\eqn\examre{
I = t^{\n-\r} B(\n, \m-\n+\r) F(\m,\n;\m+\r;1-t)
}
where $B(\n, \m-\n+\r)$ is a Beta function.

Here we would like to reproduce the result by using the Mellin 
transformation technique of \mmf. For the moment we first assume 
$Re(\r) > 0$. Note that from \dmmd,
\eqn\mmi{
(x+t)^{-\r} = {1 \ov \Ga (\r)}  {1 \ov 2 \pi i}
\int_{c-i \infty}^{c+ i \infty} \! ds \,
\Ga(\r + s) \Ga(-s) \, t^s \, x^{-\r -s} 
}
where $-Re(\r) < c <0$. 
Plugging the above expression into \exam, we get,
\eqn\mmg{\eqalign{
I & = {1 \ov \Ga (\r)}  {1 \ov 2 \pi i}
\int_{c-i \infty}^{c+ i \infty} \! ds \,
\Ga(\r + s) \Ga(-s)  \, t^{s} 
\int_0^{\infty} \! dx \, \, x^{\n -\r - s -1} (1 + x)^{-\m} \cr
& = {1 \ov \Ga (\r) \Ga (\m) } {1 \ov 2 \pi i}
\int_{c-i \infty}^{c+ i \infty} \! ds \, t^{s} 
\Ga(\r + s) \, \Ga(-s)  \,  \Ga(\n -\r -s ) \, \Ga(\m + \r -\n +s) 
}}
and the convergence of $x$-integral requires:
\eqn\mmh{
-Re(\m + \r - \n) < Re(s) < Re(\n - \r)
}
Since  $x$-integral in \mmg\ has generated new pole
sequences in the complex $s$-plane, we have to check that 
they won't cause any ambiguity in carrying the inverse-Mellin 
integral in the second line of \mmg.

\fig{Pole structures and contours. 
a: $Re(\m)> Re(\n) >Re(\r)>0$;
b: $Re(\r)> Re(\n) >Re(\m)>0$.
}{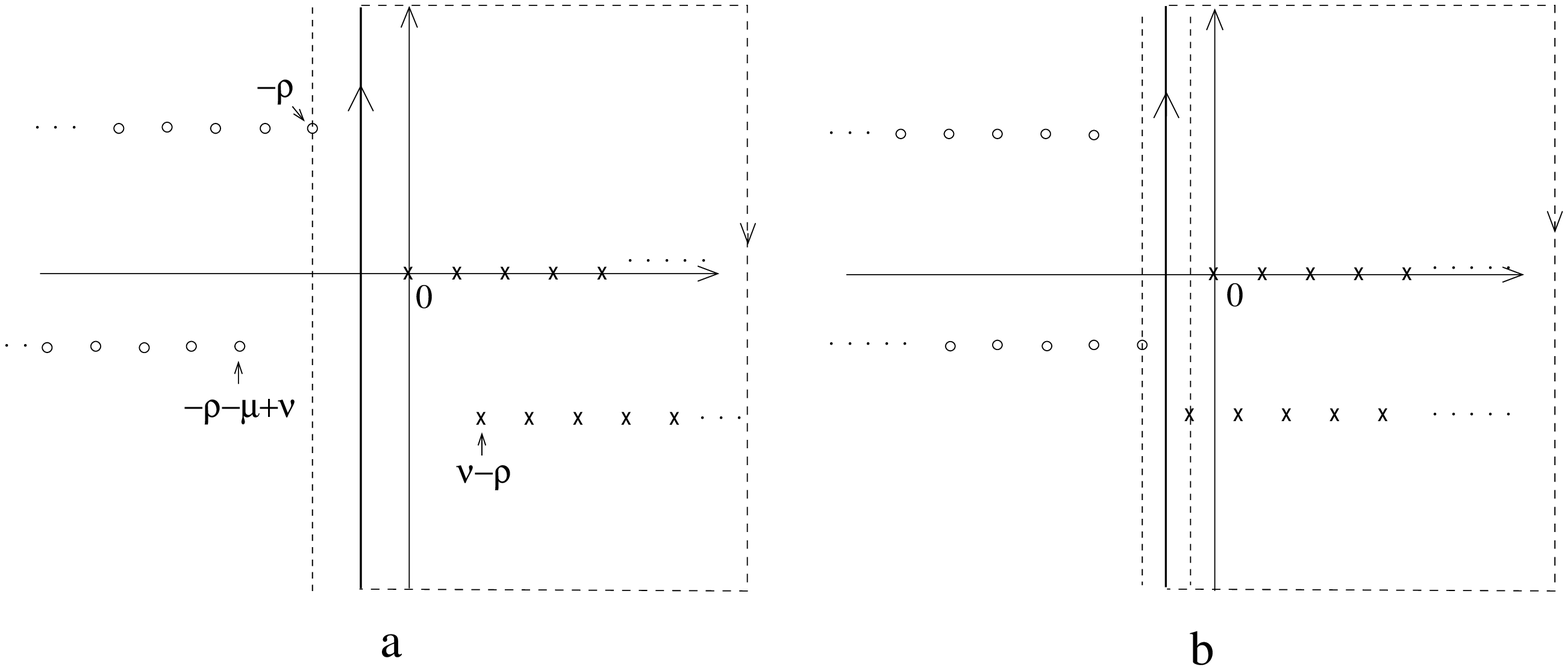}{12.8 truecm}
\figlabel\polei

\fig{Pole structure and analytic continuation.
a: $Re(\r)> Re(\n) >0$, $Re(\m)<0$; b: general values of parameters.
}{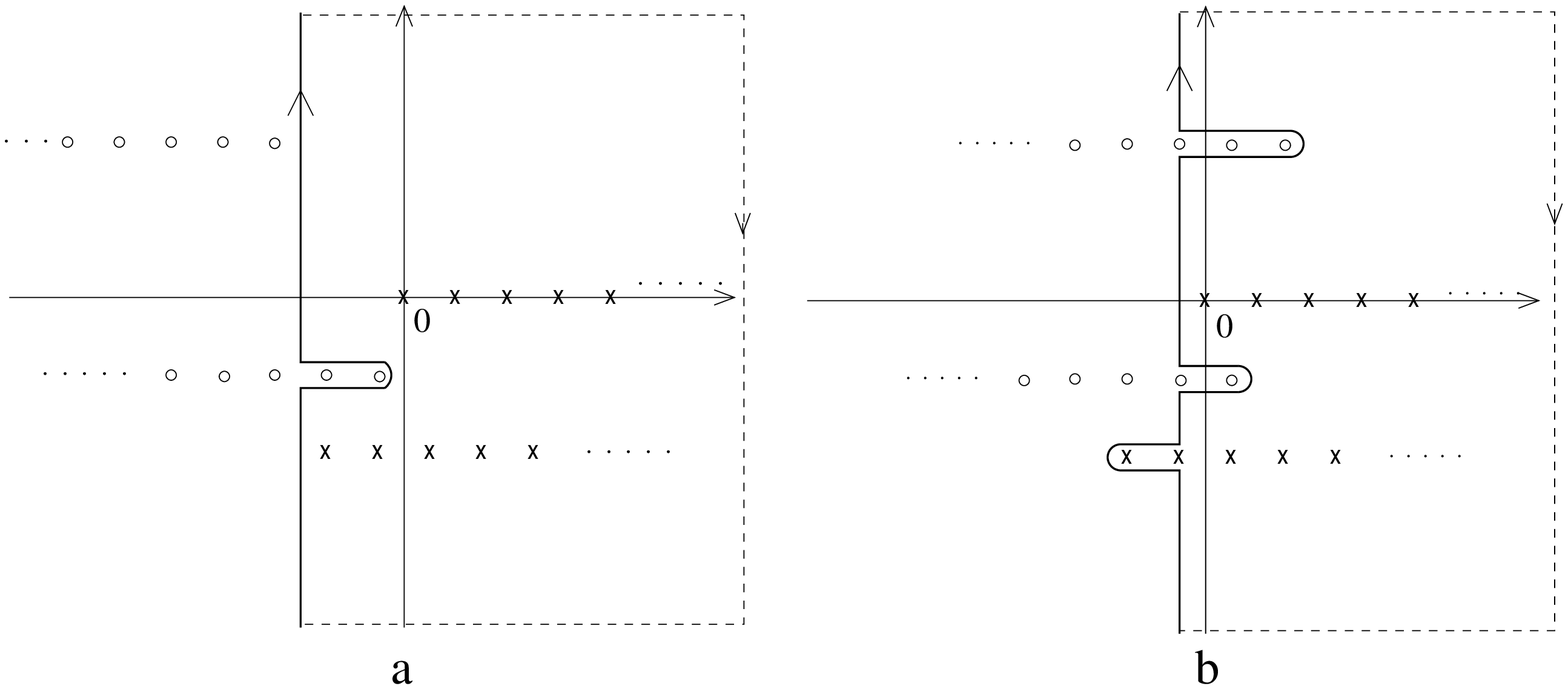}{12.8 truecm}
\figlabel\polea

\subsec{$Re(\m)> Re(\n) >Re(\r)>0$}
In this case, it is easy to check that there is no overlap 
between descending and ascending pole sequences and 
the new pole sequences sit outside the strip $-Re(\r) < c <0$
(which means the convergence condition \mmh\ is trivially satisfied), 
where the  Inverse Mellin transform \mmi\ is defined.
Thus there is no ambiguity in defining the integral in \mmg\
and we can take the integral around a contour $C$, which
consists of the path in \mmg\ and encloses the right half plane.
See figure \polei a for the pole structure and the contour.
Since $|t|<1$ the contribution from part of the contour 
other than \mmg\  vanishes and \mmg\ can be written by the calculus
of residues as the sum of the residues of the integrand at the poles 
$s=0,1, \cdots$ and $s= \n - \r + n, n=0,1, \cdots$. 
It is easy to see the sum  gives us \examre.

\subsec{$Re(\m), Re(\n), Re(\r) >0$}
In this case, there is still  no overlap 
between descending and ascending pole sequences, but there 
are poles sitting inside the strip  $-Re(\r) < c <0$ if 
$\m < \n$ or $\n < \r$, which seems to cause ambiguity 
in the choice of $c$ in \mmi\
as they may enclose different poles inside the strip.
However, the convergent condition \mmh\ requires that
we squeeze the integration path in \mmg\ into a smaller strip 
${\rm Max}[-Re(\m + \r -\n), -Re(\r)]<c< {\rm Min}[0, Re(\n-\r)] $. 
It is clear that in this refined strip there is indeed no ambiguity to 
define the integral and again we get the desired result. 
See figure \polei b for the pole structure and the contour for the case:
$Re(\r)> Re(\n) >Re(\m)>0$
\subsec{$Re(\m)<0, Re(\n), Re(\r) >0$ and others}
Now the convergent condition \mmh\ can no longer be 
satisfied. There is an overlap between the ascending poles  
from $\Ga(\n-\r-s)$ and descending poles from $\Ga(\m+\r-\n+s)$ 
and there does not exist a uniform strip that the inverse Mellin 
integral \mmg\
is well defined. In this case we can define the integral by analytic 
continuation
from the convergent region of $\m$. It is clear that as we vary $\m$ 
continuously from
$\m>0$ to $\m<0$, the only way to avoid the sudden jump of the value of 
integral 
by crossing the poles from $\Ga(\m + \r - \n+s)$ 
is to deform the integration path so that it still separates 
the descending and ascending pole sequences (see figure \polea a).

It is obvious that by repeating the above 
procedure of deforming the path of \mmg\ (see figure \polea b) we can 
analytically continue the integral \exam\ to arbitrary complex values of $\m,\n,\r$
except for some discrete surfaces in the space of  $\m,\n,\r$ where one or more
of $\r, \m$  and $\n, \m+\r-\n$ become non-positive integers.
In these cases there are coincidences between the ascending poles and 
descending 
poles and it is no longer possible to separate them. 
The analytic continuation breaks 
down at these surfaces. The pathology at 
$\m, \r = -k, \,\, k =0,1,2,\cdots$ 
may  be attributed to the method we are using (see eq \mmi)\foot{When 
$\m, \r = -k, \,\, k =0,1,2,\cdots$, 
$(1+x)^{-\m}$ or $(x+t)^{-\r}$ becomes  
finite series and can be expanded 
directly to evaluate the integral. The result 
can be expressed in terms of the terminating 
series of hypergeometric functions.},
while at $\n,  \m+\r-\n =-k, \,\, k =0,1,2,\cdots$ the analytic
continuation truly breaks down (similar to the poles in Gamma functions).

\appendix{B} {Detailed evaluation of $J$}

Here we present the details of the calculation leading from \jflag\
to \jans. To avoid making formulas too long, we will 
suppress the prefactors (numerical constants and powers of $x_{ij}$) 
of the integrals and give their final expression only  at the
end. We use the following 
definitions,
\eqn\notations{\eqalign{
& \td{\l}  = d - \l,\,\,\,
\n = \l - {d \ov 2}, \,\,\,
\n_{i} = \l_i - {d \ov 2}, \,\,
i = 1, \cdots, 4,  \cr
& \l_{ij} = \l_i + \l_j, \,\,\, 
\De_{ij} = \l_i - \l_j, \,\,\,
\ep_{ij} = \l_{ij} - \l,\,\,\,
\te_{ij} = \l_{ij} - \td{\l}, \cr
& \d_{1} = \l + \De_{12},\,\,\,
\d_{2} = \l + \De_{21},\,\,\,
\d_{3} = \l + \De_{34},\,\,\,
\d_{4} = \l + \De_{43},\,\,\, \cr
& \tde_1 = \td{\l} + \De_{12},\,\,\,
\tde_2 = \td{\l} + \De_{21},\,\,\,
\tde_3 = \td{\l} + \De_{34},\,\,\,
\tde_4 = \td{\l} + \De_{43}.\,\,\,
}}

$J(s)$ in eq \jflag\ can be further simplified 
by applying the inversion trick \free:
set $x_4=0$, then use a simultaneous inversion of the external coordinates
and the integration variables, $u_\m  \ra u_\m / |u|^2, \, 
v_\m \ra v_\m / |v|^2$, 
$\vec{x}_i = \vec{x}'_i / | \vec{x}'_i|^2, \,\,\, i=1,2,3$, 
after which $J$ becomes,  
$$
J(s) = {2^{\l + 2s} \ov |x_1|^{2 \l_1} |x_2|^{2 \l_2}
|x_3|^{2 \l_3}} \,\, \int \! {d u_0 d^d u \over u_0^{d+1}}
{d v_0 d^d v \over v_0^{d+1}} \ 
 ({u_0 \over |u-x'_1|^2})^{\l_1} ({u_0 \over |u-x'_2|^2})^{\l_2} 
$$
\eqn\jlag{
\times \,\,
({u_0 v_0 \ov u_0^2 + v_0^2 + |\vec{u}-\vec{v}|^2 })^{\l + 2s}
({v_0 \over |v-x'_3|^2})^{\l_3} v_0^{\l_4} \ .
}

Using
\eqn\schw{
{\Ga(\l) \ov z^{\l}} = \int_0^\infty \! d \r \, \r^{\l-1} e^{-\r z}
}
we can rewrite equation \jlag\ as, 
$$
J(s) = \int_{0}^{\infty} \! d \r_1 d \r_2  d \r_3  d \r \,\, \r_1^{\l_1-1} 
\r_2^{\l_2-1} \r_3^{\l_3-1} \r^{\l + 2s-1}
\int du_0 d \vec{u} dv_0 d \vec{v} \,\, u_0^{\td{\ep}_{12} + 2s -1} \,\,
v_0^{\te_{34} + 2s -1}  
$$
$$
\times \,\, 
e^{- (\r_1 + \r_2 + \r) u_0^2} \,
e^{-(\r + \r_3) v_0^2} \, 
\exp{\{- \r_1 |\vec{u} - \vec{x}'_1|^2 - 
\r_2 |\vec{u} - \vec{x}'_2|^2 - \r |\vec{u} - \vec{v}_2|^2 -
\r_3 |\vec{v} - \vec{x}'_3|^2 \} }
$$
Integrating over $u_0, v_0$ we get \foot{The convergence of $u_0$,
$v_0$ integrals require $Re(\td{\ep}_{12} + 2s)>0$ and 
$Re(\td{\ep}_{34} + 2s)>0$. Since $Re(s) \sim 0$, the convergence 
conditions are indeed satisfied with $\l_i, \l > d/2$.},
$$
J(s) =   \int_{0}^{\infty} \! d \r_1 d \r_2  d \r_3  d \r \,\,
\r_1^{\l_1-1} \r_2^{\l_2-1} \r_3^{\l_3-1} \r^{\l + 2s - 1}
({1 \ov \r_1 + \r_2 + \r})^{{\te_{12} \over 2} + s}
({1 \ov  \r_3 + \r})^{{\te_{34} \over 2} + s}  
$$
$$
\times \,\, 
\int \! d \vec{u} d \vec{v} \, \,
\exp{\{- \r_1 |\vec{u} - \vec{x}'_1|^2 - 
\r_2 |\vec{u} - \vec{x}'_2|^2 - \r |\vec{u} - \vec{v}|^2 -
\r_3 |\vec{v} - \vec{x}'_3|^2 \} }
$$
Now we use the following expression
\eqn\liu{
\int d^d \vec{u} \exp { \{ - \sum_{i} \r_i |\vec{u} - \vec{x}_i|^2 \} }
= ({\pi \ov  \sum_{i} \r_i} )^{d \ov 2} 
\exp { \{ - { \sum_{i<j} \r_i \r_j x_{ij}^2 \ov 
\sum_{i} \r_i } \} }
}
to integrate over $\vec{u}, \vec{v}$, which leads to,
$$
J(s) = \int_{0}^{\infty} \! 
d \r_1 d \r_2  d \r_3  d \r \,\,
\r_1^{\l_1-1} \r_2^{\l_2-1} \r_3^{\l_3-1} \r^{\l + 2s - 1}
({1 \ov \r_1 + \r_2 + \r})^{{\te_{12} \over 2} + s}
({1 \ov  \r_3 + \r})^{{\te_{34} \over 2} + s}  \times 
$$
$$
({1 \ov \r_3 (\r_1 + \r_2 + \r) + \r (\r_1 + \r_2)})^{{d \over 2}} \,
\exp{\{- {\r \r_2 \r_3 |x'_{23}|^2 + \r \r_1 \r_3 |x'_{13}|^2 
+ \r_1 \r_2 (\r_3 + \r) |x'_{12}|^2 \ov  \r_3 (\r_1 + \r_2 + \r) 
+ \r (\r_1 + \r_2)}\} }
$$
where $|x'_{ij}| = |\vec{x}'_i - \vec{x}'_j|$. 

Let $\r_i \ra  \r \r_i, \, i=1,2,3$ and integrate over $\r$,
$$
J(s) = \int_{0}^{\infty} \!
d \r_1 d \r_2  d \r_3 \,\,
\r_1^{\l_1-1} \r_2^{\l_2-1} \r_3^{\l_3-1} 
({1 \ov 1 + \r_1 + \r_2})^{{\te_{12} \over 2} + s}
({1 \ov  1+ \r_3 })^{{\te_{34} \over 2} + s}
\times
$$
$$ 
[ (\r_1 + \r_2) (1 + \r_3) + \r_3 ]^{{\l_{12} + \De_{34} -d \ov 2}} \,
[{1 \ov  \r_2 \r_3 |x'_{23}|^2 + \r_1 \r_3 |x'_{13}|^2 
+ \r_1 \r_2 (\r_3 + 1) |x'_{12}|^2}]^{{\l_{12} + \De_{34} \ov 2}}
$$
Note that the convergence of $\r$-integral requires $\l_{12} + \De_{34} >0$.

Now define new variables, $\r_1 = \s u, \, \r_2 = \s (1-u), \r_3 = \r $,
and 
\eqn\pcror{
\eta = {|x'_{13}|^2 \ov |x'_{12}|^2}, \,\,\,\,
\x = {|x'_{23}|^2 \ov  |x'_{12}|^2}, \,\,\,\,
z= {\e \ov 1-u} + {\x \ov u},
}
after which the integrals become,
$$
J(s) = \int_{0}^{1} du u^{{\De_{12} - \De_{34} \ov 2} - 1}
(1-u)^{{\De_{21} - \De_{34} \ov 2} - 1}
\int d \s d \r \,\, 
\s^{{\l_{12}-\De_{34} \ov 2}-1}
(1+\s)^{-{\te_{12} \over 2}-s} 
$$
$$
\times \,\, 
\r^{\l_3 -1}(1 + \r)^{-{\te_{34} \over 2}-s}
(\s + \r + \s \r)^{{\l_{12} + \De_{34} - d \ov 2}}
(\s (1 + \r) + \r z)^{-{\l_{12} + \De_{34} \ov 2}}
$$
Further  define $t = {\r \ov 1 + \r}$, so that
$$
J = \int_{0}^{1} \! du \, u^{{\De_{12} - \De_{34} \ov 2} - 1}
(1-u)^{{\De_{21} - \De_{34} \ov 2} - 1}
\int_0^{\infty} \!  d \s \int_0^1 \! dt \,\,
 \s^{{\l_{12}-\De_{34} \ov 2}-1}
(1 + \s)^{-{\te_{12} \over 2}-s}
$$
\eqn\jint{
\times \,\, 
t^{\l_3 -1} (1-t)^{{\d_4 \ov 2} +s -1}
(\s + t)^{{\l_{12} + \De_{34} - d \ov 2}} 
(\s + tz)^{-{\l_{12} + \De_{34} \ov 2}}
}

Our next step is to use the inverse Mellin transformation eq \dmmd,
\eqn\mellin{
({1 \ov 1+x})^{\a} = {1 \over \Ga(\a)} {1 \ov 2 \pi i}
\int_{-i \infty}^{i \infty} \! ds \, \Ga(-s) \, \Ga(\a + s)  x^s
}
in  $(\s + tz)^{-{\l_{12} + \De_{34} \ov 2}}$ in eq \jint.
Then the $\s$-$t$ part of the integrals in eq \jint\ becomes
\eqn\interm{
{1 \ov 2 \pi i}
\int_{-i \infty}^{i \infty} \! d s_1 \, \, \Ga(-s_1) \,
\Ga({\l_{12} + \De_{34} \ov 2} + s_1) 
\, \, z^{-s_1 - {\l_{12} + \De_{34} \ov 2}} \, J_1
}
with
$$
J_1 = \int_0^{\infty} \!  d \s \!
\int_0^1 \!  dt \,\,
t^{{\l_{34}-\l_{12} \ov 2}- s_1 -1} (1-t)^{{\d_4 \ov 2} +s -1} 
\s^{{\l_{12}-\De_{34} \ov 2}+ s_1 -1} 
(1 + \s)^{-{\te_{12} \over 2}-s}
(\s + t)^{{\l_{12} + \De_{34} - d \ov 2}} 
$$

After $\s$-integration we get,
$$
J_1 = B(s_1 + {\l_{12}-\De_{34} \ov 2}, s-s_1 - {\ep_{12} \ov 2})\,\,
\times
$$
$$
\int_0^1 \! dt \,\,  t^{{\d_4 \ov 2} +s -1}
(1-t)^{{\l_{34}+\l_{12} -d \ov 2} -1} \,\, 
F({\te_{12} \over 2}+s, s_1 + {\l_{12}-\De_{34} \ov 2}; s+ {\d_4 \ov 2},t).
$$
Notice that the power in $t$ coincides with the third parameter of 
the hypergeometric function inside the integral. 
In this case the integration over $t$ can be done very easily and we get
\eqn\iii{
\
J_1 ={ \Ga ({\l_{34}+\l_{12} -d \ov 2}) \ov \Ga (\l_4)}
 { \Ga(s_1 + {\l_{12}-\De_{34} \ov 2}) \Ga ( s-s_1 - {\ep_{12} \ov 2}) 
\Ga ({\l_{34} - \l_{12} \ov 2} - s_1) \ov \Ga (s-s_1 + {\te_{34} \over 2})}
\
}

Plugging \iii\ back into \interm, \jint\  and rearranging the integrals
$$
J(s) =C_2 \, {1 \ov 2 \pi i}
\int_{-i \infty}^{i \infty} \! d s_1 \, \, 
\Ga(-s_1) \, 
\Ga({\l_{12} + \De_{34} \ov 2} + s_1) \,
\Ga(s_1 + {\l_{12}-\De_{34} \ov 2}) \,
\Ga ({\l_{34} - \l_{12} \ov 2} - s_1) \,
$$
\eqn\iiij{
\times \,\, 
{\Ga ( s-s_1 - {\ep_{12} \ov 2}) \ov \Ga (s-s_1 + {\te_{34} \over 2})}\,\,
\int_{0}^{1} du u^{{\De_{12} - \De_{34} \ov 2} - 1}
(1-u)^{{\De_{21} - \De_{34} \ov 2} - 1} 
({\e \ov 1-u} + {\x \ov u})^{-s_1 - {\l_{12} + \De_{34} \ov 2}} \,
}
The integral in the second line gives us a 
hypergeometric function and the final 
expression for $J$ is (we shifted $s_1$ by $s_1+\l_{12} \ra s_1$),
$$
J(s) =C_2 \, {1 \ov 2 \pi i}
\int_{{\cal C}}
\! d s_1 \,  \x^{-s_1 - {\De_{34} \ov 2}} \,
\Ga({\l_{12} \ov 2} -s_1) \Ga ({\l_{34} \ov 2} - s_1) \,
F( {\De_{34} \ov 2} +s_1 ,{\De_{12} \ov 2} + s_1; 2s_1; 1-{\e \ov \x})\,
\times
$$
\eqn\jans{
{\Ga( {\De_{34} \ov 2} + s_1) \Ga({\De_{43} \ov 2} + s_1)
\Ga( {\De_{12} \ov 2} + s_1) \Ga({\De_{21} \ov 2} + s_1) 
\ov  \Ga (2s_1)} \,
{\Ga ({\l \ov 2} +  s-s_1 ) 
 \ov \Ga ({\l_{12} + \te_{34} \over 2} + s-s_1)} \,
}

Now restore $x_4$ by taking $x_i \ra x_i-x_4, \,i=1,2,3 $.
$\e$ and $\x$ defined in \pcror\ become the  cross ratios,
\eqn\cror{
\eta = {|\vec{x}'_{14} -\vec{x}'_{34}|^2 \ov 
|\vec{x}'_{14} -\vec{x}'_{24}|^2} = {|x_{13}|^2 |x_{24}|^2
\ov  |x_{12}|^2 |x_{34}|^2}, \,\,\,\,\,\,\,\,\,\,\,\,
\x = {|\vec{x}'_{24} -\vec{x}'_{34}|^2 \ov 
|\vec{x}'_{14} -\vec{x}'_{24}|^2} = {|x_{14}|^2 |x_{23}|^2
\ov  |x_{12}|^2 |x_{34}|^2}
}
Finally, the prefactor $C_2$ is given by,
$$ 
C_2 = {\pi^d \ov 4} 
{ \Ga ({\l_{12} + \l_{34} -d \ov 2}) 
\Ga (s + {\te_{12} \ov 2}) \Ga (s + {\te_{34} \ov 2})
\ov \Ga(\l_1)\Ga(\l_2) \Ga(\l_3) \Ga(\l_4) \Ga(\l + 2s)}
{2^{\l + 2s} \ov  |x_{12}|^{\l_{12} + \De_{34}}
|x_{14}|^{\De_{12}-\De_{34}}  
|x_{24}|^{\De_{21}- \De_{34}}
|x_{34}|^{2 \l_3}}
$$

We note that it can be checked that the integrals in intermediate steps 
from \jlag\  to \jans\ are  convergent only when the parameters 
satisfy the following conditions (${\rm Re}(s) \sim 0$),
\eqn\fcon{
{\d_1 \ov 2},{\d_2 \ov 2},{\d_3 \ov 2},{\d_4 \ov 2}  >0, \,\,\, \,
{\l_{12} \pm \De_{34} \ov 2}>0, \,\,\, \,
{\l_{34} \pm \De_{12} \ov 2}>0
}
For the above range of parameters there is no overlap between ascending 
and descending poles in \jans\ and the $s_1$-integral can be 
unambiguously defined by squeezing (also determined by the convergence 
of the intermediate integrals) the integration path ${\cal C}$ to 
lie inside the strip:
$$
{\rm Max}[ {\De_{12} \ov 2},{\De_{21} \ov 2},{\De_{34} \ov 2}, 
{\De_{34} \ov 2}]
< {\rm Re}(s_1) < {\rm Min} [{\l \ov 2}, {\l_{12} \ov 2}, {\l_{34} \ov 2}]
$$
When  parameters are outside the range of \fcon,  some 
integrals in intermediate steps may not be convergent and 
can only be defined by analytic continuation. Further there are 
overlaps between the ascending and descending pole sequences in \jans\
and there does not exist a uniform strip in which  the Mellin integral 
can be well defined. In this case we can define the integral in \jans\ by 
analytic continuation by deforming  the 
integration path ${\cal C}$ so that it  separates the ascending and 
descending poles. In  Appendix A we have given  a detailed 
discussion of this  procedure in a simpler example. 
The whole discussion can be applied to this
more complicated case without change.
We note that when some poles from ascending and descending series 
coincide with one another, e.g. if  one or more  
of $ {\l_{12} \pm \De_{34} \ov 2}$, 
${\l_{34} \pm \De_{12} \ov 2}$,
${\d_i \ov 2}, i=1, \cdots, 4$ are non-positive integers,  there is no
way to separate the ascending and descending poles.  
The analytic  continuation breaks down at these  points. 

\appendix{C} {Evaluation of contact contribution}

Here we would like evaluate eq \da. Again we will suppress the prefactor 
of the integrals most of the time and follow the notations defined 
in \notations.

As in Appendix B, using eq \schw\ and integrating over $u_0$, we get,
\eqn\db{\eqalign{
S_{c} & =  
\int \! \Pi_{i=1}^4 \, d \r_i \, \r_i^{\l_i-1} 
{1 \ov (\r_1 + \r_2 + \r_3 + \r_4)^{{\l_{12} + \l_{34} -d \ov 2}}}
\int \! d \vec{u} \, \exp{ [-\sum_i \r_i |\vec{u}-\vec{x}_i|^2 ] } \cr
& = 
\int \! \Pi_{i=1}^4 \, d \r_i \, \r_i^{\l_i-1}
{1 \ov (\r_1 + \r_2 + \r_3 + \r_4)^{{\l_{12}+ \l_{34} \ov 2}}}
\exp {\{- { \sum_{i<j=1}^{4} \r_i \r_j |x_{ij}|^2 \ov 
 \r_1 + \r_2 + \r_3 + \r_4 } \} }
}}
where in the second line we have used eq \liu.
Now  let 
$\r'_i = \r_i (\r_1 + \r_2 + \r_3 + \r_4)^{-\ha}$
and note that $det({\del \r_i \ov \del \r'_j}) = 2 (\sum_{i=1}^{4}
\r'_i)^4$, 
\eqn\db{
S_{c} = {1 \ov  \pi^{3d/2}} {\Ga({\l_{12} + \l_{34} -d \ov 2})
\ov \Ga(\n_1)\Ga(\n_2)\Ga(\n_3)\Ga(\n_4)}
K(\l_1, \l_2, \l_3,\l_4)
}
where we have included the numerical prefactor and 
$K$ is defined by, 
\eqn\kkk{
K(\l_1,\l_2, \l_3,\l_4) = 
\int_0^\infty \!  d \r_1 d \r_2  d \r_3  d \r_4 \,\,
\Pi_i \r_i^{\l_i -1 }
\exp{ \big[ - \sum_{i<j=1}^{4} \r_i \r_j |x_{ij}|^2 \big ] }.
}

Since $K$ has translational invariance, we can take $x_4 =0$. 
Defining $\r'_i = \r_i |x_i|^2, \,\, i=1,2,3$,  we find that,
$$
\sum_{i<j=1}^{4} \r_i \r_j |x_{ij}|^2 
= \r_4 ( \r'_1 + \r'_2 + \r'_3) + \sum_{i<j=1}^{4} \r'_i \r'_j |x'_{ij}|^2
$$
where $\vec{x}'_{ij} = \vec{x}'_i -  \vec{x}'_j$
and $\vec{x}'_i = {\vec{x}_i \ov |x_i|^2}$. Then in terms 
of $\r'_i$ (we omit primes on $\r_i$ below) and $x'_i$, $J$ becomes,
\eqn\aa{\eqalign{
K & = {1 \ov |x_1|^{2 \l_1}|x_2|^{2 \l_2} |x_3|^{2 \l_3}}
\int_{0}^{\infty} \! \Pi_{i=1}^{3} d \r_i \r_i^{\l_i -1} 
\exp{ \big[ - \r_4 ( \r_1 + \r_2 + \r_3) -
\sum_{i<j=1}^{3} \r_i \r_j |x'_{ij}|^2 \big] } \cr
& ={\Ga(\l_4) \ov |x_1|^{2 \l_1} |x_2|^{2 \l_2}|x_3|^{2 \l_3}} 
\int_{0}^{\infty} \! \Pi_{i=1}^{3} d \r_i \r_i^{\l_i -1} 
{1 \ov (\r_1 + \r_2 + \r_3 )^{\l_4}} 
\exp{ \big[ -
\sum_{i<j=1}^{3} \r_i \r_j |x'_{ij}|^2 \big] }
}}

Now let $\r_3 = \a$, $\r_1 = \a \b$, $\r_2 = \a \g$,
\eqn\ab{\eqalign{
K & = \int \! d \a d \b d \g \,\, {\b^{\l_1 -1} \g^{\l_2 -1}
\a^{\l_{12}+ \De_{34}  -1} \ov ( 1 + \b + \g)^{\l_4}} 
\exp{ [ - \a^2 (\b |x'_{13}|^{2} + \g |x'_{23}|^{2} 
+ \b \g |x'_{12}|^{2} )]} \cr
& = 
\int \! d \b d \g {\b^{\l_1 -1} \g^{\l_2 -1}
\ov ( 1 + \b + \g)^{\l_4}} \,
{ 1 \ov [\b |x'_{13}|^{2} + \g |x'_{23}|^{2} 
+ \b \g |x'_{12}|^{2} ]^{{\l_{12}+ \De_{34} \ov 2}}}
}}
Further let $\b = \s u$, $\g= \s (1-u)$ and 
define, 
$$
\e = {|x'_{13}|^{2} \ov |x'_{12}|^{2}},\,\,\,\,
\x ={|x'_{23}|^{2} \ov |x'_{12}|^{2}}, \,\,\,\,
z= {\e \ov 1-u} + {\x \ov u},
$$ 
after which $K$ becomes,
\eqn\ac{
K =  
\int_{0}^{1} \! du u^{{\De_{12}- \De_{34} \ov 2}-1} 
(1-u)^{{\De_{21}- \De_{34} \ov 2}-1} \, \int_{0}^{\infty} \! d \s 
\s^{{\l_{12} - \De_{34} \ov 2} -1} (1 + \s)^{-\l_4}
(\s + z)^{- {\l_{12}+ \De_{34} \ov 2}} \
}
$\s$-integral above is nothing but the familiar integral representation 
of a hypergeometric function. 

We now restore $x_4$, after which $\x$ and $\e$ become the cross ratios 
defined in \cror. Including the prefactor of the integral, we get
the final expression for $K$,
$$
K = \ha { \Ga (\l_3) \Ga (\l_4) 
\Ga({\l_{12}- \De_{34} \ov 2})\Ga({\l_{12} + \De_{34} \ov 2})
\ov \Ga({\l_{12} + \l_{34} \ov 2})}
{1 \ov |x_{12}|^{\l_{12} + \De_{34}}
|x_{14}|^{\De_{12}-\De_{34}}  
|x_{24}|^{\De_{21}- \De_{34}}
|x_{34}|^{2 \l_3}} \times 
$$
\eqn\af{
\int_{0}^{1} \! du u^{{\De_{12}- \De_{34} \ov 2}-1} 
(1-u)^{{\De_{21}- \De_{34} \ov 2}-1} 
z^{-\De_{34}}
F({\l_{12}+ \De_{34} \ov 2}, 
{\l_{12}- \De_{34} \ov 2}; {\l_{12}+ \l_{34} \ov 2};1-{1 \ov z}).
}

Alternatively we may use the  
Mellin-Barnes representation for a hypergeometric function,
$$
F(a,b;c;1-z) = {\Ga(c) \ov \Ga(a) \Ga(b) \Ga(c-a) \Ga(c-b)}
{1 \ov 2 \pi i} \int_{-i \infty}^{i \infty} \! ds z^s 
\Ga(-s) \Ga(c-a-b-s) \Ga(a+s) \Ga(b+s)
$$
in eq \af, and then integrate over $u$. In this form $S_c$ can be 
written as, 
$$
S_c  = C_c \, {1 \ov 2 \pi i} 
\int_{\l_{12}-i \infty}^{\l_{12}+ i \infty} \!
 d s \, \x^{-s} \,\,
 \Ga({\l_{12} \ov 2} -s) 
\Ga({\l_{34}\ov 2} - s) \,
F( {\De_{34} \ov 2} +s ,{\De_{12} \ov 2} + s; 2s; 1-{\e \ov \x}) 
\times
$$
\eqn\ae{
{\Ga( {\De_{34} \ov 2} + s) \Ga({\De_{43} \ov 2} + s)
\Ga( {\De_{12} \ov 2} + s) \Ga({\De_{21} \ov 2} + s) 
\ov  \Ga (2s)}
}
with 
$$
C_c = {1 \ov 2 \pi^{3d/2}} {\Ga({\l_{12} + \l_{34} -d \ov 2})
\ov \Ga(\n_1)\Ga(\n_2)\Ga(\n_3)\Ga(\n_4)} 
{1 \ov |x_{12}|^{\l_{12}} \, |x_{14}|^{\De_{12}} \,
|x_{24}|^{\De_{21} - \De_{34}} \, |x_{23}|^{\De_{34}} \, 
|x_{34}|^{ \l_{34}}}
$$

\bigskip\bigskip

\centerline {\ \bf Acknowledgements}
\bigskip
I would like to thank V. Balasubramanian, T. Banks, G. Chalmers, M. Douglas,
D. Freedman, I. Klebanov, A. Lawrence, M. Li, J. Maldacena,
S. Mathur, A. Matusis, S. Minwalla,
A. Rajaraman,  L. Rastelli, W. Siegel, A. Tseytlin, 
E. Witten for discussions, especially, 
A. Tseytlin for a careful reading of the manuscript and E. Witten for 
suggesting the interpretation in terms of double-trace operators
and possible explanation of logarithms. 
I am also grateful to Aspen Center for Physics and theoretical 
physics groups at  Rutgers, MIT, Princeton  
and SUNY at Stony Brook for hospitality during  various stages
of the work. This work was supported in part by PPARC.

\listrefs
\bye